\documentclass[%
 aip,
 amsmath,amssymb,
 reprint,
]{revtex4-1}

\usepackage{graphicx}% Include figure files
\usepackage{dcolumn}% Align table columns on decimal point
\usepackage{bm}% bold math
\usepackage[utf8]{inputenc}
\usepackage[T1]{fontenc}
\usepackage{mathptmx}
\usepackage{etoolbox}
\usepackage{color}
\DeclareUnicodeCharacter{2212}{-}
\DeclareUnicodeCharacter{0301}{-}
\DeclareUnicodeCharacter{03B5}{-}
\makeatletter
\def\@email#1#2{%
 \endgroup
 \patchcmd{\titleblock@produce}
  {\frontmatter@RRAPformat}
  {\frontmatter@RRAPformat{\produce@RRAP{*#1\href{mailto:#2}{#2}}}\frontmatter@RRAPformat}
  {}{}
}%
\makeatother
\begin{document}

\preprint{AIP/123-QED}

\title{Effect of substrate mismatch, orientation, and flexibility on heterogeneous ice nucleation}
% Force line breaks with \\

\author{M. Camarillo}
\affiliation{Departamento de Química Física, Facultad de Ciencias Químicas, Universidad Complutense de Madrid, 28040 Madrid, Spain}

\author{J. Oller-Iscar}
\affiliation{Department of Chemical Engineering, Universidad Polit\'{e}cnica de Madrid, Jos\'{e} Guti\'{e}rrez Abascal 2, 28006, Madrid, Spain}

\author{M.M. Conde}
\affiliation{Department of Chemical Engineering, Universidad Polit\'{e}cnica de Madrid, Jos\'{e} Guti\'{e}rrez Abascal 2, 28006, Madrid, Spain}

\author{Jorge Ramírez}
\affiliation{Department of Chemical Engineering, Universidad Polit\'{e}cnica de Madrid, Jos\'{e} Guti\'{e}rrez Abascal 2, 28006, Madrid, Spain}

\author{E. Sanz*}
\affiliation{Departamento de Química Física, Facultad de Ciencias Químicas, Universidad Complutense de Madrid, 28040 Madrid, Spain}
\email{esa01@ucm.es}

\date{\today}% It is always \today, today,

\begin{abstract}
Heterogeneous nucleation is the main path to ice formation on Earth.
The ice nucleating ability of a certain substrate is mainly determined by
both molecular interactions and the structural mismatch between the ice and the 
substrate lattices. 
We focus on the latter factor using molecular simulations of the mW model. 
Quantifying the effect of structural mismatch alone is challenging due to its coupling 
with molecular interactions.
To disentangle both factors, we use a
substrate composed of water molecules in such a way that any variation on 
the nucleation temperature can be exclusively ascribed to the structural mismatch.
We find that a one per cent increase of structural mismatch leads to a decrease of approximately 4 K 
in the nucleation temperature. 
We also analyse the effect of the orientation of the substrate with respect to the liquid. 
\textcolor{black}{The three main ice orientations (basal, primary prism and secondary prism) have a similar ice nucleating ability.}
We finally asses the effect of lattice flexibility by comparing substrates where molecules are 
immobile with others where a certain freedom to fluctuate around the lattice positions is allowed. 
Interestingly, we find that the latter type of substrate is more efficient in nucleating ice because
it can adapt its structure to that of ice.  

\end{abstract}

\maketitle

\section{\label{sec:intro}Introduction}

Heterogeneous ice nucleation refers to the process by which ice crystals form on surfaces of foreign particles that act as ice-nucleating agents. It is a phenomenon of pivotal importance because it constitutes the main mechanism of ice formation 
on Earth \cite{sanz2013homogeneous}.
The study of heterogeneous ice nucleation is essential for a wide range of fields, including atmospheric science, climate research, biology, cryobiology, and various industrial applications. Advancements in understanding this phenomenon can lead to improved weather forecasting, climate models, cryopreservation techniques, aviation safety measures or 
artificial snow production \cite{hoose2012heterogeneous}.

There is a wide variety of ice nucleating agents \cite{murray2012ice,kanji2017overview,hoose2010classical,harrison2016not}
such as mineral dust, soot, biological aerosols or organic compounds that act as sites for ice crystal formation in clouds. 
Many experimental studies have been undertaken to understand and quantify the ice-nucleating ability of these
and other particles \cite{murray2012ice,hoose2012heterogeneous,atkinson2013importance,zhang2018control,whale2015ice,sosso2022role}.

Complementarily, models have been
developed to understand and predict experiments \cite{hoose2010classical,knopf2013water,cabriolu2015ice,khvorostyanov2000new,zobrist2008heterogeneous,liu2005ice,chen2008parameterizing,barahona2009parameterizing},
and simulations \cite{lupi2016pre,li2018surface,kiselev2017active,fraux2014note,soni2021microscopic,zielke2015molecular,zielke2016simulations,soni2021unraveling,pedevilla2018heterogeneous,cox2013microscopic} are used to unveil key molecular 
features of ice-nucleating particles.
The most common approach is to use realistic potentials for water and for the selected substrate under investigation. \cite{kiselev2017active,fraux2014note,soni2021microscopic,zielke2015molecular,zielke2016simulations,soni2021unraveling,pedevilla2018heterogeneous,cox2013microscopic,sosso2016ice,glatz2016surface,sosso2018unravelling,sosso2022role,pedevilla2016can}.
This strategy enables making predictions directly relevant to experiments \cite{kiselev2017active}, although it
requires a reliable substrate-water interaction potential, which is an aspect that has not been 
carefully taken into consideration. 

In this work we trade realism of the substrate for the possibility of shedding light on 
fundamental questions regarding heterogeneous ice nucleation.
In particular, our main goal is to unravel the effect of the mismatch between
the ice and the substrate lattices, an issue 
raised by Turnbull in 1952 \cite{turnbull1952nucleation}.
The difficulty of answering this question lies on the fact that the influence of the mismatch  
is usually coupled to that of substrate-water interactions
\cite{bi2017enhanced,fitzner2015many,glatz2018heterogeneous,cox2015molecularI,cox2015molecularII,valeriani2022deep,soni2021microscopic,reinhardt2014effects,li2017roles,zhang2018control,lupi2014does,lu2021effect,cox2013microscopic,fitzner2020predicting}.
To circumvent this problem we use substrates composed of water molecules, a strategy 
inspired by previous works \cite{mithen2014epitaxial,mithen2014computer,reinhardt2014effects}.
Our approach is similar in spirit to that taken
by Mithen et al. \cite{mithen2014epitaxial,mithen2014computer} where the nucleation
ability is probed as a function of the lattice mismatch for a substrate and a liquid both composed of
Lennard-Jones particles. Reinhardt et al. \cite{reinhardt2014effects}
also used this strategy for water, although they mainly focused on the role of interactions and did not
perform any quantitative analysis on that of lattice mismatch. In this work, we quantify ice 
nucleating ability by computing the temperature
at which heterogeneous nucleation occurs at a certain rate for substrates with different 
structural mismatches with ice. 
We find that the nucleation temperature approximately decreases 4 K when the mismatch between 
ice and the substrate increases by 1 per cent. 

Moreover, we interrogate other fundamental aspects of heterogeneous nucleation 
such as the effect of lattice flexibility or that of the orientation of the substrate. 
We analyse the former by comparing the ability to induce ice nucleation of substrates with fixed molecular
positions with that of solids composed of molecules that can wander around their lattice sites. 
In accordance with studies using realistic substrate potentials \cite{zielke2016simulations}, we find that flexibility 
favours nucleation via a mechanism in which stress is shared between the substrate and the 
emerging ice nucleus. 
Regarding the lattice orientation effect, we find that the three main ice directions, basal, primary prism (pI) and secondary prism (pII)
have a very similar ability to induce ice nucleation, being pI and basal the most and the 
least efficient nucleants, respectively, by a narrow margin.

Overall, our work leaves aside the complexity found in real systems \cite{wilson2003ice} 
to tackle fundamental questions regarding heterogeneous ice nucleation such as 
the effects of lattice mismatch, orientation and flexibility.

%\date{\today}
%

\section{\label{sec:meth}Methodology}
We use the mW water model \cite{molinero2009water}, whose melting temperature is 273 K \cite{hudait2016free}.
We run our simulations with the Large-scale
Atomic/Molecular Massively Parallel Simulator (LAMMPS) Molecular Dynamics package \cite{plimpton1995fast} in the 
NVT ensemble. Temperature is kept constant with the Nosé-Hoover 
thermostat \cite{nose1984unified,hoover1985canonical}.
We integrate the equations
of motion using the velocity-Verlet integrator with a
3 fs time step. 

\begin{table}[!hbt]
    \centering
    \begin{tabular}{c c c c c c}
         $L_x$ && $L_y$ && $L_z$  \\
         \hline
         $8.853$ && $7.671$  && $7.203$ \\
         \hline
         \hline
         x&&y&&z\\
         \hline

            0.000 && 0.000 && 0.000 \\
            0.500 && 0.000	&& 0.000 \\
            0.250 && 0.500 && 0.000 \\
            0.750 && 0.500 && 0.000 \\
            0.250 && 0.167 && 0.125 \\
            0.750 && 0.167 && 0.125 \\
            0.000 && 0.667 && 0.125 \\
            0.500 && 0.667 && 0.125 \\
            0.250 && 0.167 && 0.500 \\
            0.750 && 0.167 && 0.500 \\
            0.000 && 0.667	&& 0.500 \\
            0.500 && 0.667 && 0.500 \\
            0.000 && 0.000 && 0.625 \\
            0.500 && 0.000 && 0.625 \\
            0.250 && 0.500 && 0.625 \\
            0.750 && 0.500 && 0.625 \\

          &&  &&  \\

    \end{tabular}
    \caption{Sides (L$_x$, L$_y$  and L$_z$) in \AA~ and particle coordinates in lattice units (x, y, z) of the ice Ih orthorhombic unit cell equilibrated at 
coexistence conditions, i. e. 1 bar and 273 K, we use to build our systems.}
    \label{tab:unitcell}
\end{table}

\begin{center}
\begin{figure}[!hbt] \centering
    \centering
    \includegraphics[clip,scale=0.135,angle=0.0]{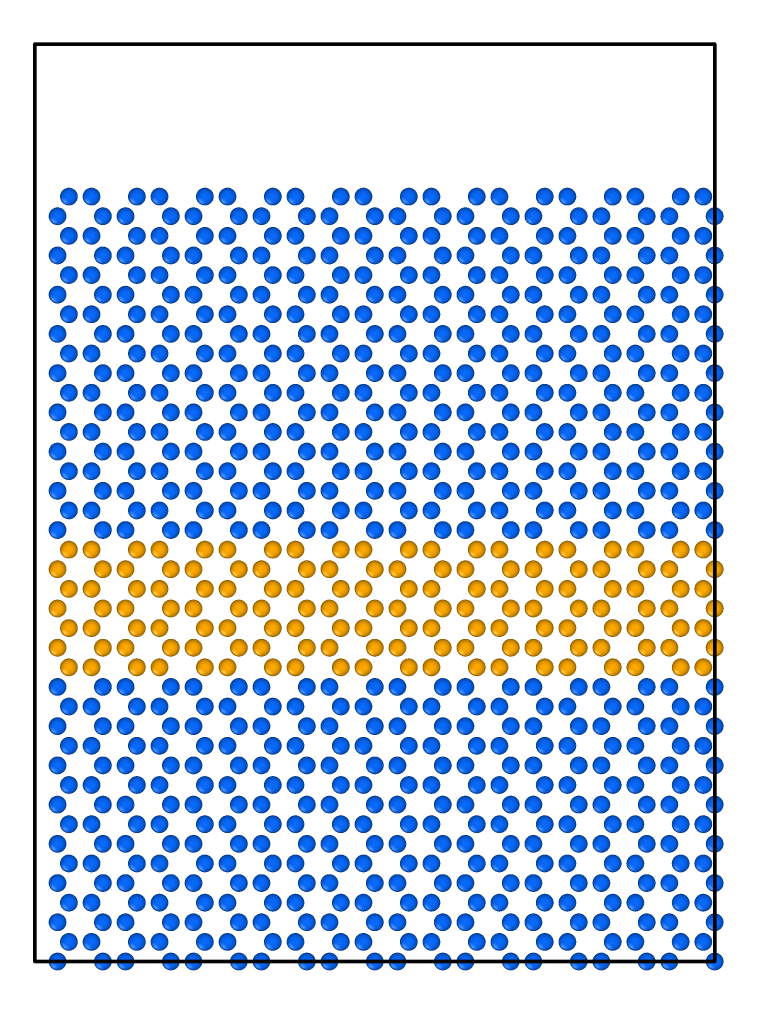}
    \includegraphics[clip,scale=0.135,angle=0.0]{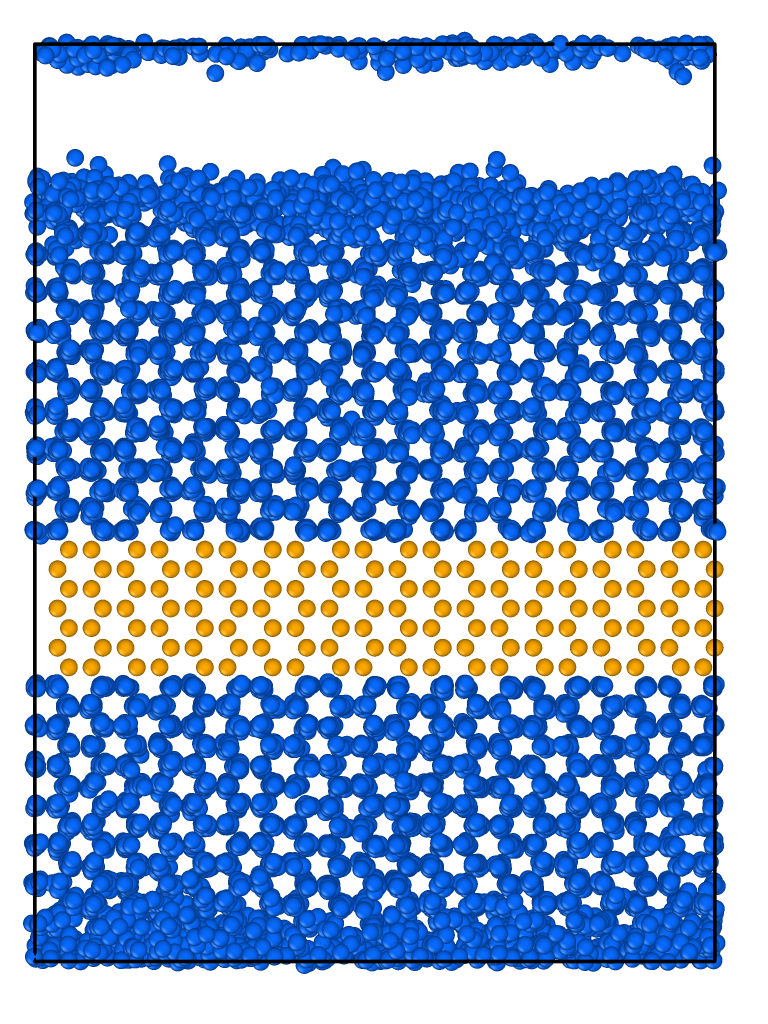}
    \includegraphics[clip,scale=0.135,angle=0.0]{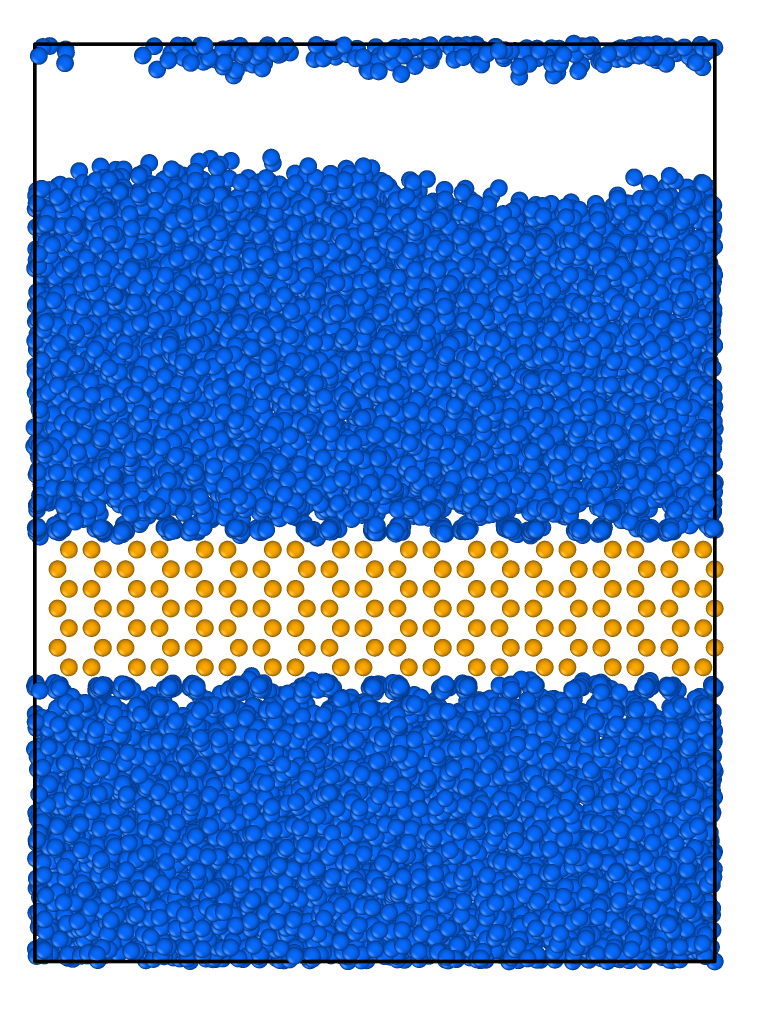}
    \caption{Sequence of snapshots illustrating our procedure to generate an initial configuration. Left: Snapshot of a side view of a replicated 
    (and stretched for $\delta$ >0) unit cell in contact with \textcolor{black}{a 15\AA~ thick} vacuum. Yellow particles correspond to the substrate \textcolor{black}{(exposing the pII plane to the liquid)} and blue particles will be melted to give rise to the initial configuration. Middle: At 300 K, a liquid layer is readily formed in the face of the replicated ice lattice in contact with vacuum. Right: At 300 K and in the NVT ensemble 
    the liquid layer propagates up to the substrate surface giving rise to the initial configuration.}
    \label{fig:snapshots-init}
\end{figure}
\end{center}

We generate the substrate coordinates by stretching/\textcolor{black}{compressing} 
an orthorhombic unit cell containing 16 molecules whose lattice parameters
correspond to the solid at coexistence conditions (i. e. 273 K and 1 bar). 
We give in table \ref{tab:unitcell} the sides of such unit cell and the coordinates of all 
particles in lattice units. 
An expanded/\textcolor{black}{compressed} unit cell is obtained by multiplying the three directions of the equilibrium unit cell at 273 K by a factor $f$.
By replicating the resulting unit cell we obtain the coordinates of a deformed ice
see Fig. \ref{fig:snapshots-init} (left)) whose percent mismatch, $\delta$, is 
given by:
\begin{equation}
 \delta = 100 \cdot |(f-1)|.   
 \label{eq:delta}
\end{equation}
A central slab of such solid, the substrate region,
is kept frozen (orange particles in Fig. \ref{fig:snapshots-init})
and the rest is simulated at 300 K in the NVT ensemble.
We apply 
periodic boundary conditions leaving an empty space along the direction perpendicular to the substrate
to melt the outermost layer \cite{vega2006absence,slater2019surface} (Fig. \ref{fig:snapshots-init}(middle)). The molten region propagates up to the frozen substrate giving rise to an initial 
configuration (Fig. \ref{fig:snapshots-init}(right)).

The mismatch parameter, $\delta$, is always referred to the unit cell at coexistence conditions. 
One may consider using the stretching/\textcolor{black}{compressing} factor with respect to the equilibrium unit cell at the temperature
of interest instead of coexistence. This would be impractical
because it would force us to build an initial configuration for every temperature.
However, the error made by considering the deformation with respect to the unit cell at coexistence is very
small: the ice density only increases by 0.2 per cent from coexistence to the 
lowest studied temperature, which means that the edges of the unit cell decrease by a factor of 1.00085 
($\delta$ = $-0.085$). With such a small volume variation of the solid with temperature 
it is definitely  not worth considering different unit cells for different temperatures.

We work with two different types of substrate, both composed 
of water molecules of the same nature as those in the liquid phase: A \emph{rigid} substrate where 
water molecules are immobile in their lattice positions, and a \emph{wells} substrate where each molecule 
is given freedom to move within a potential energy well 
of 2.043  \AA~diameter and 8$k_bT$ depth.  
Details on how we
build the well-particle interaction are given in Ref. \cite{sanchez2022homogeneous}.
The comparison of rigid versus wells substrates enables assessing the role of lattice rigidity 
on ice nucleating ability. 

Due to the presence of a liquid-vapour interface
the average pressure of the system is nearly 0 bar, which is essentially equivalent
to 1 bar when dealing with condensed phases as we do in the present work.
We use NVT instead of NpT simulations for these reasons:
(i) a 0 bar pressure is perfectly maintained in the NVT ensemble
(ii)  it is easier to generate configurations in the NVT ensemble via melting at a 
free interface as previously explained (iii) we avoid altering the structure of the interface
with volume moves. 
A system as that shown in Fig.  \ref{fig:snapshots-init}(right) is first equilibrated at 300 K 
and then quenched to a temperature below melting where the simulation is run
until ice crystallization is observed. In Table \ref{tab:sistemas} we provide details on the size and 
the orientation of the substrate with respect to the liquid of the systems studied in this work with stretched substrates.
 
\begin{table}[!hbt] \centering
\caption{Specifications for the systems studied in this work with stretched substrates. 
$A_m$ is the area of the substrate face exposed to the liquid, $N_s$ is the number
of molecules in the substrate, $N_l$ is the number of molecules in the liquid, $L_s$ is the 
substrate thickness and $n_xn_yn_z$ is the number of times that the original orthorhombic ice unit cell (reported in Table 
\ref{tab:unitcell})
was replicated along $x$, $y$ and $z$ to build the initial lattice (see Fig. \ref{fig:snapshots-init} (left)). 
The substrate face that is exposed to the 
liquid is also indicated (the basal, pI and pII orientations correspond to the 
the xy, xz and yz planes, respectively).
T$_n$ is the nucleation temperature, defined as the temperature for which heterogeneous nucleation takes place at 
a rate of 10$^{23.6}$ nuclei/(m$^2$s). \textcolor{black}{The error bar in the nucleation temperature is 0.5 K.}}
\label{tab:sistemas}
  \begin{center}
    \begin{tabular}{ c c c c c c c c c c c c c c c c c}
\hline
\hline
\multicolumn{15}{c}{Rigid substrate} \\
$\delta$ && $A_m$/(nm$^2$) && $N_s$ && $N_l$ && $L_s$ && $n_xn_yn_z$ && Plane && T$_n$ / K\\
          
\hline
  5 && 137.1 && 5400 && 30600 && 13.943 && 10x15x15 && pII && 254.6 \\
  7 && 142.3 && 5400 && 30600 && 14.208 && 10x15x15 && pII && 244.4 \\
  8 && 145.0 && 5400 && 30600 && 14.341 && 10x15x15 && pII && 241.4 \\
% 10 && 150.4 && 5400 && 30600 && 14.607 && 10x15x15 && pII && 234.0 \\
\hline
\multicolumn{15}{c}{Wells substrate} \\
  5 && 137.1 && 5400 && 30600 && 9.295 && 10x15x15 && pII && 255.8 \\
  7 && 142.3 && 5400 && 30600 && 9.472 && 10x15x15 && pII && 245.1 \\
  8 && 145.0 && 5400 && 30600 && 9.561 && 10x15x15 && pII && 242.9 \\
\hline
\hline
  5 && 168.5 && 5400 && 30600 && 10.399 && 15x15x10 && basal && 255.5 \\
  5 && 158.3 && 5400 && 30600 && 10.737 && 15x10x15 && pI && 257.0 \\
  5 && 137.1 && 5400 && 30600 && 9.295 && 10x15x15 && pII && 256.0 \\
\hline
  7 && 174.9 && 5400 && 30600 && 10.597 && 15x15x10 && basal && 245.5 \\
  7 && 164.3 && 5400 && 30600 && 8.208 && 15x10x15 && pI && 247.0 \\
  7 && 142.3 && 5400 && 30600 && 9.472 && 10x15x15 && pII &&  246.0 \\
\hline
  8 && 178.2 && 5400 && 30600 && 7.779 && 15x15x10 && basal && 243.0 \\
  8 && 167.3 && 5400 && 30600 && 8.285 && 15x10x15 && pI && 244.0 \\
  8 && 145.0 && 5400 && 30600 && 9.561 && 10x15x15 && pII && 243.5 \\
\hline
\hline
    \end{tabular}
  \end{center}
\end{table}

The heterogeneous nucleation rate is defined as the number of crystal clusters that appear and grow per unit time and unit of substrate area. 
We use the following expression to 
obtain the nucleation rate from spontaneous nucleation events \cite{filion2010crystal,espinosa2019heterogeneous}:
\begin{equation}
    J=\frac{1}{2At_{ind}}    
    \label{eq:rate}
\end{equation}
where $A$ is the area of the substrate in contact with the liquid (which is multiplied by 2
because there are two substrate-water interfaces as shown in Fig. \ref{fig:snapshots-init} (right)), and $t_{ind}$ is the average induction time 
required for the nucleation of an ice embryo on the substrate. The numerator of this expression is
1 because only \emph{one} nucleus appears when there is an induction 
period (that there is an induction period implies that a configuration with a critical nucleus
is unlikely and, therefore, the likelihood of having two or more critical nuclei simultaneously is negligible).
The induction time is identified by a sudden potential energy drop caused by
the quick growth of the nucleated ice crystal. 
The use of Eq. \ref{eq:rate} is limited to the temperature range where
an induction period is appreciated. 
At too high temperatures no nucleation events are observed in our simulation time whereas at too low temperatures
the induction time is zero because multiple nuclei readily appear and grow.

\textcolor{black}{We compare the ability of different substrates to induce ice nucleation
through  the nucleation temperature, T$_n$, which is defined as the temperature for which 
a certain value of the nucleation rate is achieved. 
 The question we try to answer is: how much does the temperature at which nucleation is observed change
 by modifying a certain parameter in the substrate (e. g. mismatch, lattice orientation or lattice flexibility)?  Of course, such temperature depends on the substrate-water area and on the observation time (the larger the area and the observation time, the higher the temperature at which nucleation is observed). The idea is to establish the comparison between substrates exposing the same area during the same time to the liquid. Under fixed area and observation time, the temperature at which nucleation is observed corresponds to a certain value of the nucleation rate (constant denominator in Eq. \ref{eq:rate}). 
 Fixing the area and the observation time naturally arises from the implementation of a set up to study nucleation. In simulations, for instance, typical areas and times are of the order of hundreds of nm$^2$ and hundreds of ns respectively.
 In summary, comparing the temperature at which nucleation 
 happens at certain rate, is equivalent to 
 compare the temperature at which ice formation takes place
 for a fixed substrate-water area and observation time. 
 Thus, comparing nucleation temperatures permits quantify the relative ability of substrates with different characteristics. The substrate inducing nucleation at a higher temperature (or, in other words, the substrate that causes a given rate value at a higher temperature) is the better nucleant.  Differences in nucleation temperatures between different substrates are expected to be fairly independent
 on the specific value of the rate chosen to define T$_n$. This is the case when the J(T) curves for different substrates are parallel to each other, as we will show later on. }

 \textcolor{black}{To try understand the relative ice nucleating efficiency of different substrates we perform an analysis of the molecular structure across the interface. In particular, we compute
the radial distribution function (rdf) of the particles contained in slabs of a few \AA~ thick parallel to the interface and look for similarities to the bulk ice rdf. 
A more sophisticated analysis consists in identifying the first rdf peak for several slabs and plot its position as a function of the distance to the interface.   
 This analysis evidences how the 
 structure evolves from the interior of the substrate up the bulk fluid.
 In some cases, we complement this 
 analysis with the average 
 q$_{12}$ local bond order
parameter \cite{steinhardt1983bond} of the particles contained 
 in each slab (considering for such calculation the 12 nearest neighbors of each particle, some of which may be placed in adjacent slabs).}

\section{Results}
\subsection{Effect of the mismatch}
\subsubsection{Isotropic positive $\delta$}
\begin{center}
\begin{figure}[!hbt] \centering
    \centering
    \includegraphics[clip,scale=0.2,angle=0.0]{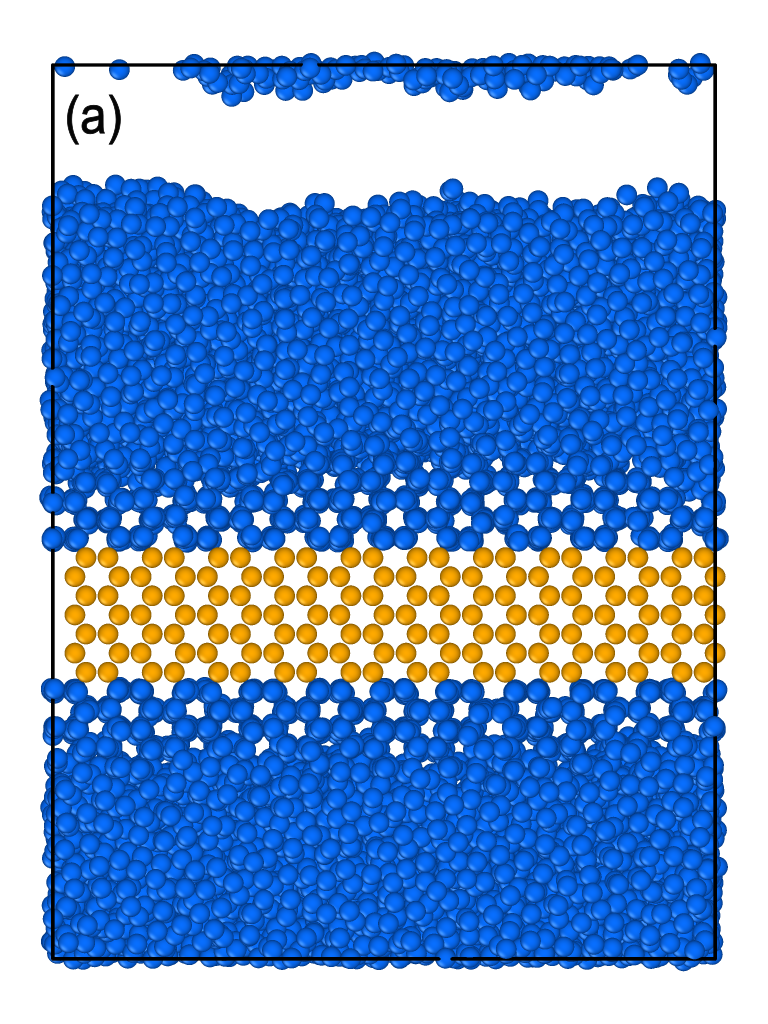} 
    \includegraphics[clip,scale=0.25,angle=0.0]{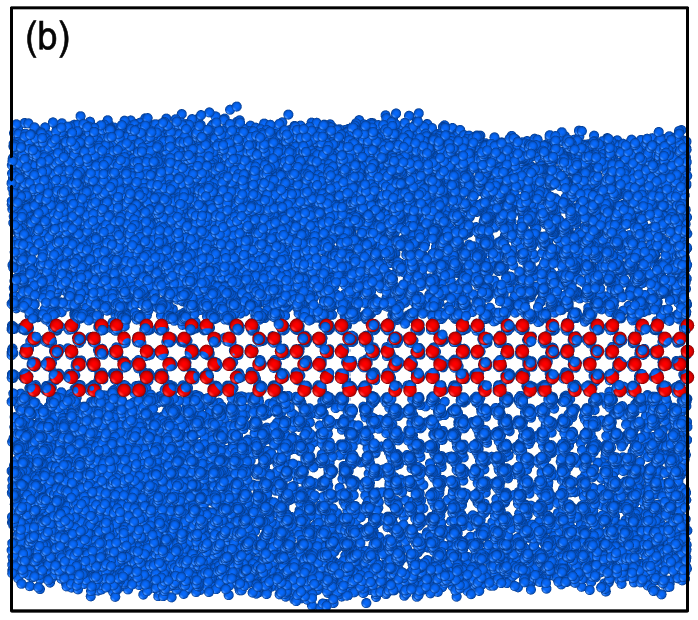}
    \caption{
    (a) Snapshot shortly after starting a simulation with a $\delta = 0$ rigid substrate at 1 K below the melting temperature. An ice layer wetting the mold is clearly seen. 
    (b) Snapshot shortly after nucleation
    on a wells substrate with $\delta = 5$  at 255.5 K.}
    \label{fig:snapshots}
\end{figure}
\end{center}

\begin{center}
\begin{figure}[!hbt] \centering
    \centering
    \includegraphics[clip,scale=0.19,angle=0.0]{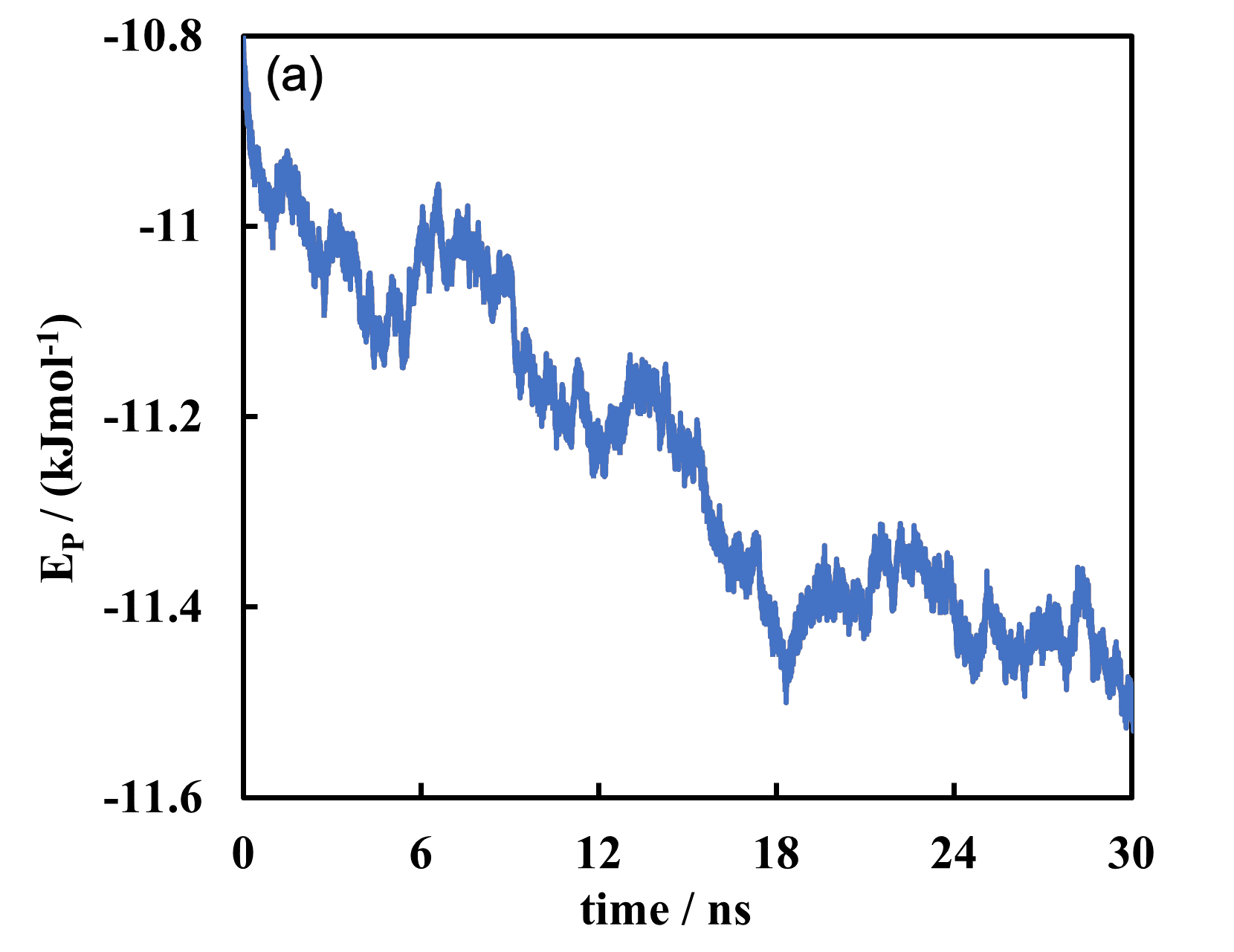}
    \includegraphics[clip,scale=0.19,angle=0.0]{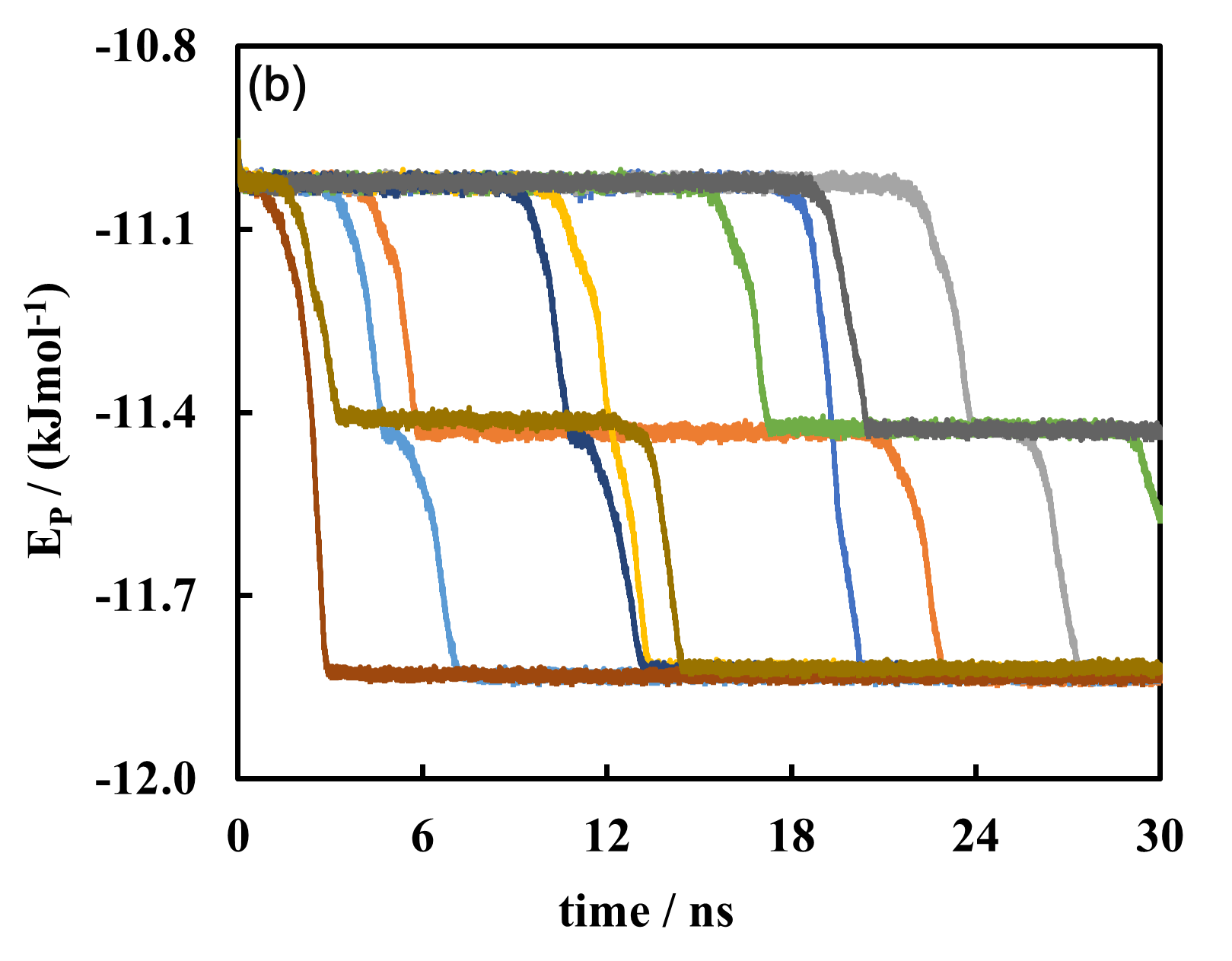}
    \includegraphics[clip,scale=0.18,angle=0.0]{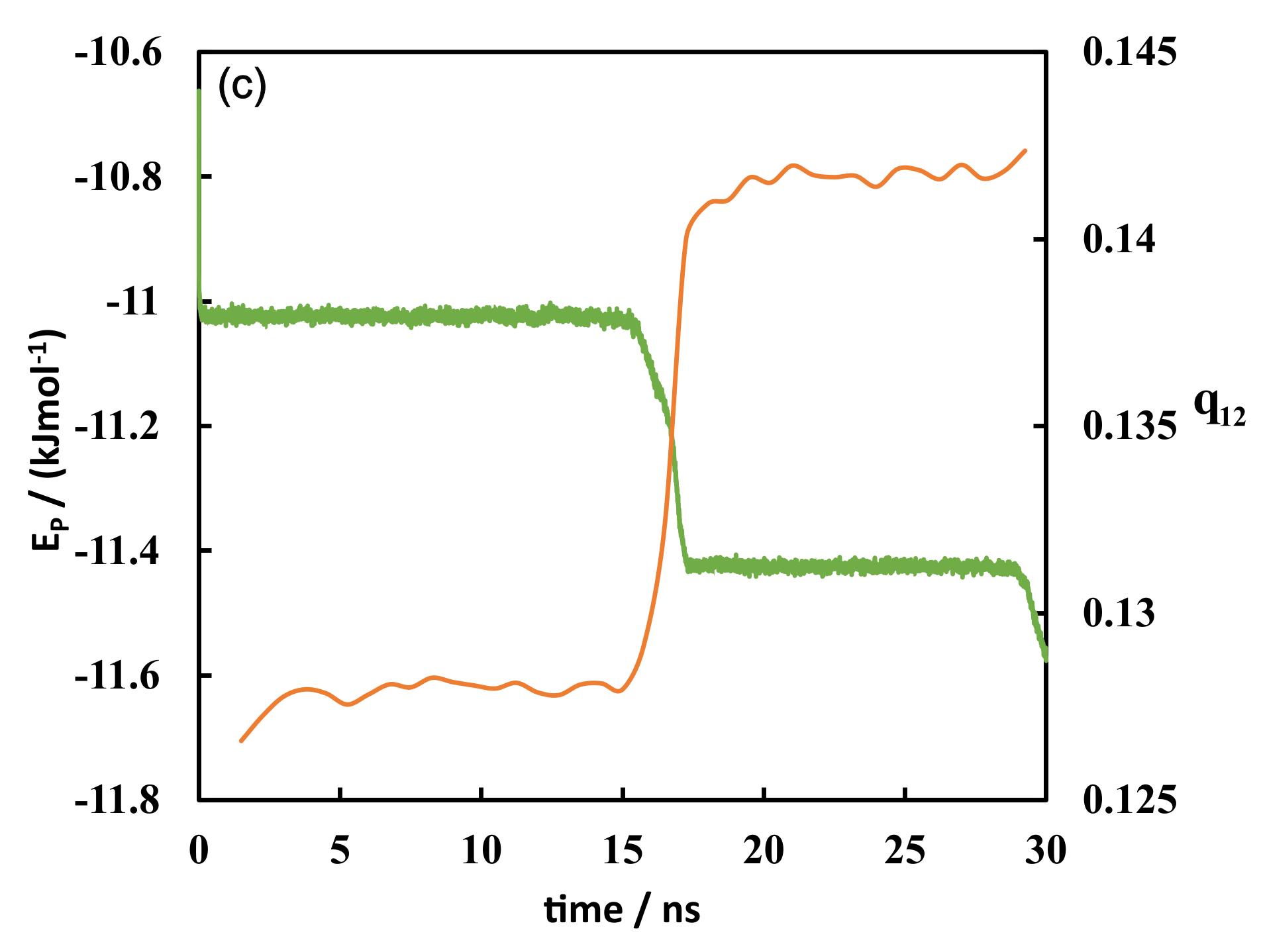}
    \caption{(a) Potential energy versus time for a trajectory at 272 K (1 K below the melting temperature) on a wells
    substrate with $\delta = 0$.
    (b) Potential energy versus time for nine trajectories of a liquid at 245.0 K on a wells substrate with $\delta = 7$.
    \textcolor{black}{(c) Time evolution of the potential energy (green) compared to that of the average q$_{12}$   local bond order parameter \cite{steinhardt1983bond} (orange) for a single nucleation trajectory of those shown in (b).}}
    \label{fig:espontaneous}
\end{figure}
\end{center}

For no mismatch between ice and the substrate, $\delta = 0$, we recover the behaviour expected for a direct coexistence
simulation \cite{ladd1977triple,garcia2006melting,conde2013determining}. At just 1 K below the melting temperature the substrate is
immediately ``wetted'' by an ice-like layer (see Fig. \ref{fig:snapshots}(a)) that subsequently grows
until the system fully crystallizes. 
The time evolution of the potential 
energy in such type of simulation is shown in Fig. \ref{fig:espontaneous}(a).
As can be seen, the potential energy continuously goes down as the liquid turns into ice.

The phenomenology radically changes for $\delta > 0$.
First of all, in order to observe ice formation, a significant supercooling must be applied
(for $\delta = 5$ a supercooling of $\sim$ 20 K is required). 
Secondly, there is no 
visible ice wetting layer percolating the simulation box. 
Instead, ice appears after some induction period via the stochastic nucleation of an ice
seed on top of the substrate as that shown in Fig. \ref{fig:snapshots}(b). 
The potential energy curves shown in Fig. \ref{fig:espontaneous}(b), corresponding to 9 different trajectories of the crystallization of mW water 
at 245.0 K on a wells substrate with $\delta = 7$, 
clearly show an initial plateau corresponding to the induction period followed by
a sudden drop, that stochastically occurs at different times for different trajectories, 
due to ice nucleation and growth. 
After the first sudden drop, the potential energy remains flat again until
there is a second abrupt drop due to ice nucleation and growth on the other side of the
substrate. 
By averaging the time at which the first sudden drop takes place
we can estimate an 
average induction time, $t_{ind}$, to obtain the heterogeneous nucleation rate via Eq. 
\ref{eq:rate}. Statistics could have been improved by also taking into account the  
second drop, but we preferred not using this information
to avoid possible finite size effects by which crystallization in one side of 
the substrate influences ice formation on the other (e. g. by a sudden release of latent heat
upon crystallization). 
\textcolor{black}{The crystallization time inferred from a potential energy drop is the same
as that obtained through an analysis of a more sophisticated local bond order parameter. This equivalence is shown in Fig. \ref{fig:espontaneous}(c), where we compare, for a nucleation trajectory, the time evolution of the potential energy with that of the q$_{12}$ 
local bond order parameter \cite{steinhardt1983bond} averaged over all particles (computed for each particle with its first 12 neighbours).}

In summary, 
whereas a mold with $\delta = 0$ causes essentially the same effect as an ice slab in contact
with the liquid and leads to crystallization just below the melting temperature, a 
substrate with $\delta > 0$ 
requires a finite supercooling to induce crystallization through a nucleation mechanism. 
Since we use substrates with the same inter-molecular interactions but a different and controlled
lattice stretch, our approach enables us to isolate and quantify
the effect of the mismatch on the ice nucleating ability of a substrate, which is the main 
target of this work. 
The idea of using substrates composed of molecules of the same nature as the 
those of the fluid is not entirely new. 
In Ref. \cite{reinhardt2014effects} 
this approach was used to investigate how the strength of hydrogen bonding affects ice nucleation, 
although the effect of mismatch alone was not systematically quantified. 
Also in Ref. \cite{mithen2014epitaxial} a similar approach as employed for the Lennard-Jones system, 
although the liquid-substrate cross interaction 
was weakened with respect to the liquid-liquid one to avoid quick crystallization. 
In both cases, the effect of lattice mismatch was quantified in a different way as compared to our work. 

\begin{center} 
\begin{figure}[!hbt] \centering
    \centering
    \includegraphics[clip,scale=0.18,angle=0.0]{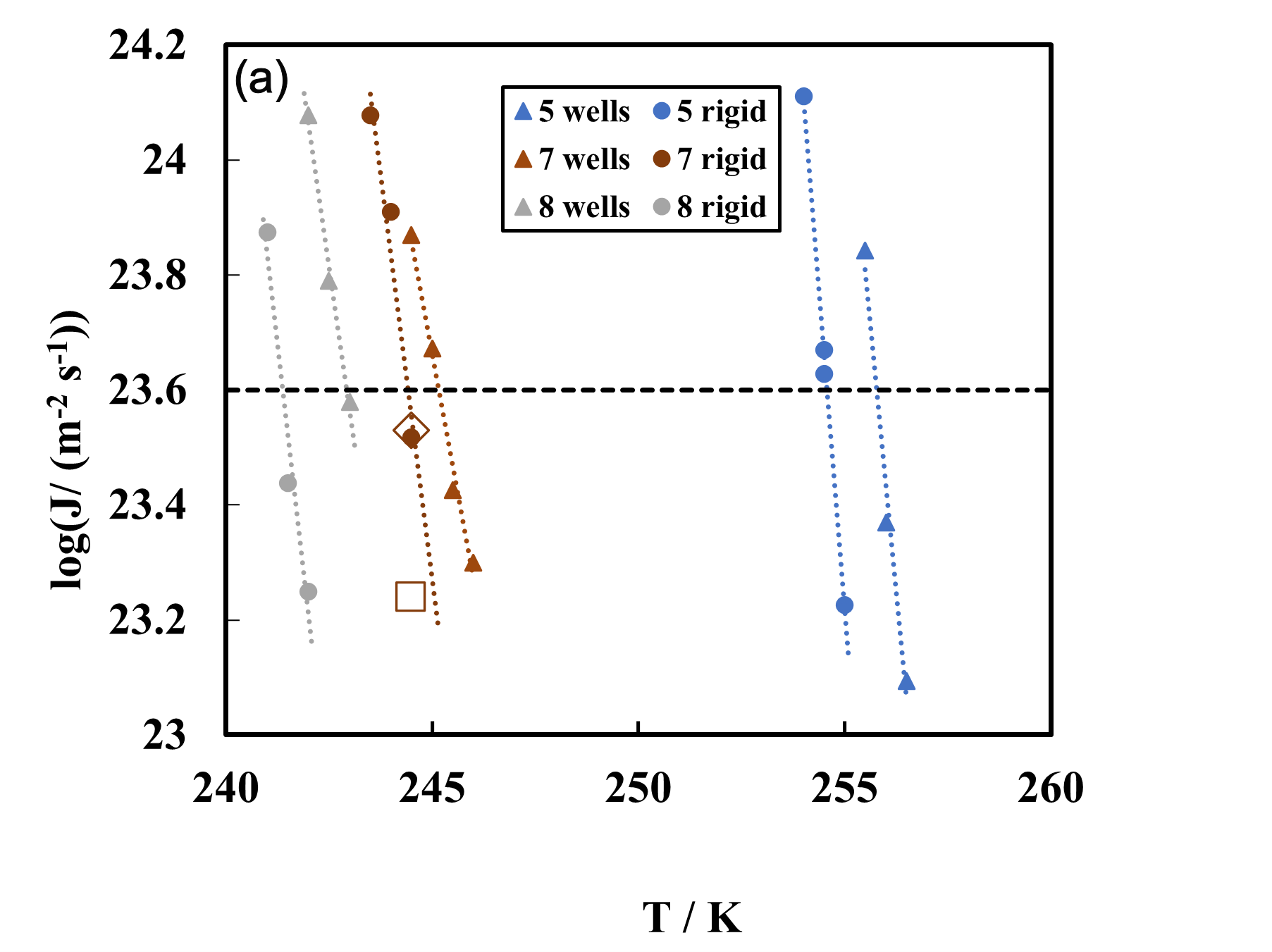}
    \includegraphics[clip,scale=0.18,angle=0.0]{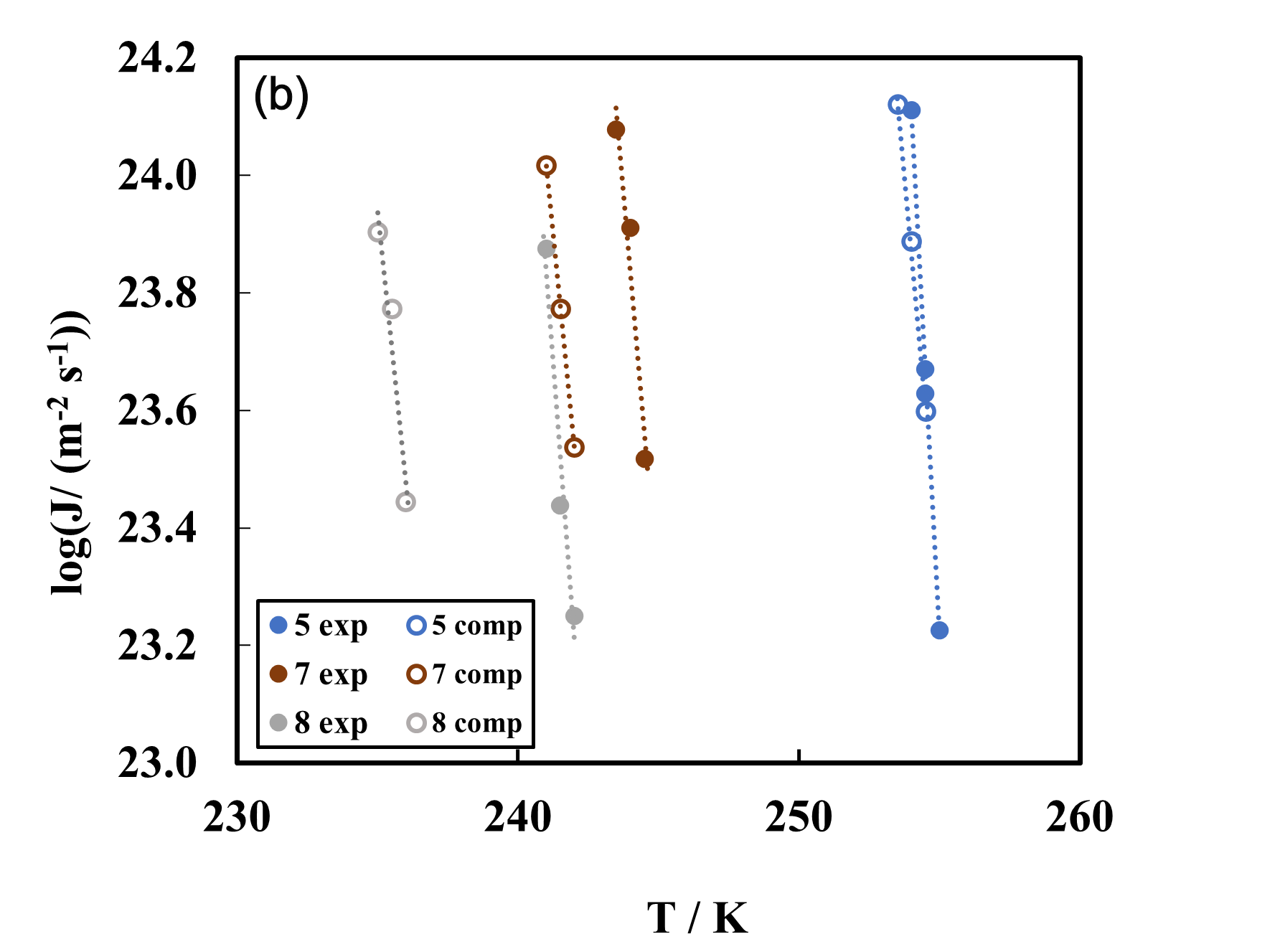}
    \caption{\textcolor{black}{(a)} Heterogeneous nucleation rate versus temperature for the different mismatches and the different
    kinds of \textcolor{black}{stretched} substrates studied in this work as indicated in the legend. These results correspond to the pII orientation 
    (exposing the yz plane of the stretched and replicated unit cell as indicated in table \ref{tab:sistemas}). The open 
    diamond (square) corresponds to a $\delta=7$ rigid substrate system having double liquid (substrate) depth than the system with which 
    the filled brown dots were obtained. \textcolor{black}{(b) Heterogeneous nucleation rate temperature dependence of
    expanded (exp) and compressed (comp) rigid substrates exposing the pII plane to the liquid.}}
    \label{fig:rate}
\end{figure}
\end{center}

The nucleation rates obtained for each of the studied molds exposing the pII plane 
to the liquid are plotted in Fig. \ref{fig:rate} 
as a function of temperature.  Triangles and circles correspond to wells and rigid molds respectively, 
whereas grey, brown and blue colors correspond to $\delta = $ 8, 7 and 5, respectively.
We have checked that the results shown in Fig. \ref{fig:rate} do not depend on 
the substrate or the liquid thicknesses. The brown square (diamond) in Fig. \ref{fig:rate}
corresponds to the rate obtained with a substrate (liquid) twice as thick as that
of the original system from which the brown circles were obtained. 
As can be seen, the results are fully consistent. 

Two comments are due in view of the curves shown in Fig. \ref{fig:rate}.
One is that, for a given $\delta$, the nucleation rate is lower for the rigid mold. The other
is that the rate curves are shifted to lower temperatures as $\delta$ increases. 
We focus first on the later effect, the dependence of the rate on $\delta$, which is the strongest one and the main 
scope of this work. 
Understandably, the supercooling needed to achieve a certain heterogeneous nucleation rate increases
with the mismatch between ice and the substrate.

To quantify to which extent the supercooling required for nucleation to occur increases 
with the mismatch we select a certain nucleation rate, $\log$(J/(m$^2$/s)) = 23.6 (given by the dotted horizontal line in 
Fig. \ref{fig:rate}), and find the associated nucleation temperature, T$_n$, for each mismatch (and substrate type) by interpolation (reported in Table \ref{tab:sistemas}). 

\textcolor{black}{
The 10$^{23.6}$ m$^{-2}$s$^{-1}$ rate has been arbitrarily chosen in the range of nucleation rates accessible by means of simulations 
\textcolor{black}{of spontaneous heterogeneous crystallization. Such range is determined by the area of the substrate and the simulation time according to Eq. \ref{eq:rate}. For an area of the order of a few hundred of squared nm and a time of tens of ns one gets log[J/(m$^3$s)] in the range of 23-24.}
The specific rate value 10$^{23.6}$ m$^{-2}$s$^{-1}$ crosses all J(T) curves shown 
in Fig. \ref{fig:rate} and enables obtaining a T$_n$($\delta$) curve by interpolation.
Despite the fact that the absolute value of T$_n$ depends on this arbitrary choice of the rate, the slope of its variation 
with the lattice mismatch is not expected to be
largely affected in view of the fact that the curves in Fig. \ref{fig:rate} are parallel to each other. Therefore, a change in the rate used to define T$_n$ would shift by the same constant
the nucleation temperature for each kind of substrate.} 

\textcolor{black}{By performing independent sets of trajectories to determine the nucleation rate we estimate the
error in log[J/(m$^3$s)] to be $\pm$ 0.1.  
We note  that, for a given mismatch, the nucleation rate is a very steep function of temperature.
As a consequence, the uncertainty in the nucleation rate does not have a strong 
impact in T$_n$. We estimate such uncertainty to be $\pm$ 0.5 K.} 

\begin{center}
\begin{figure}[!hbt] \centering
    \centering
    \includegraphics[clip,scale=0.7,angle=0.0]{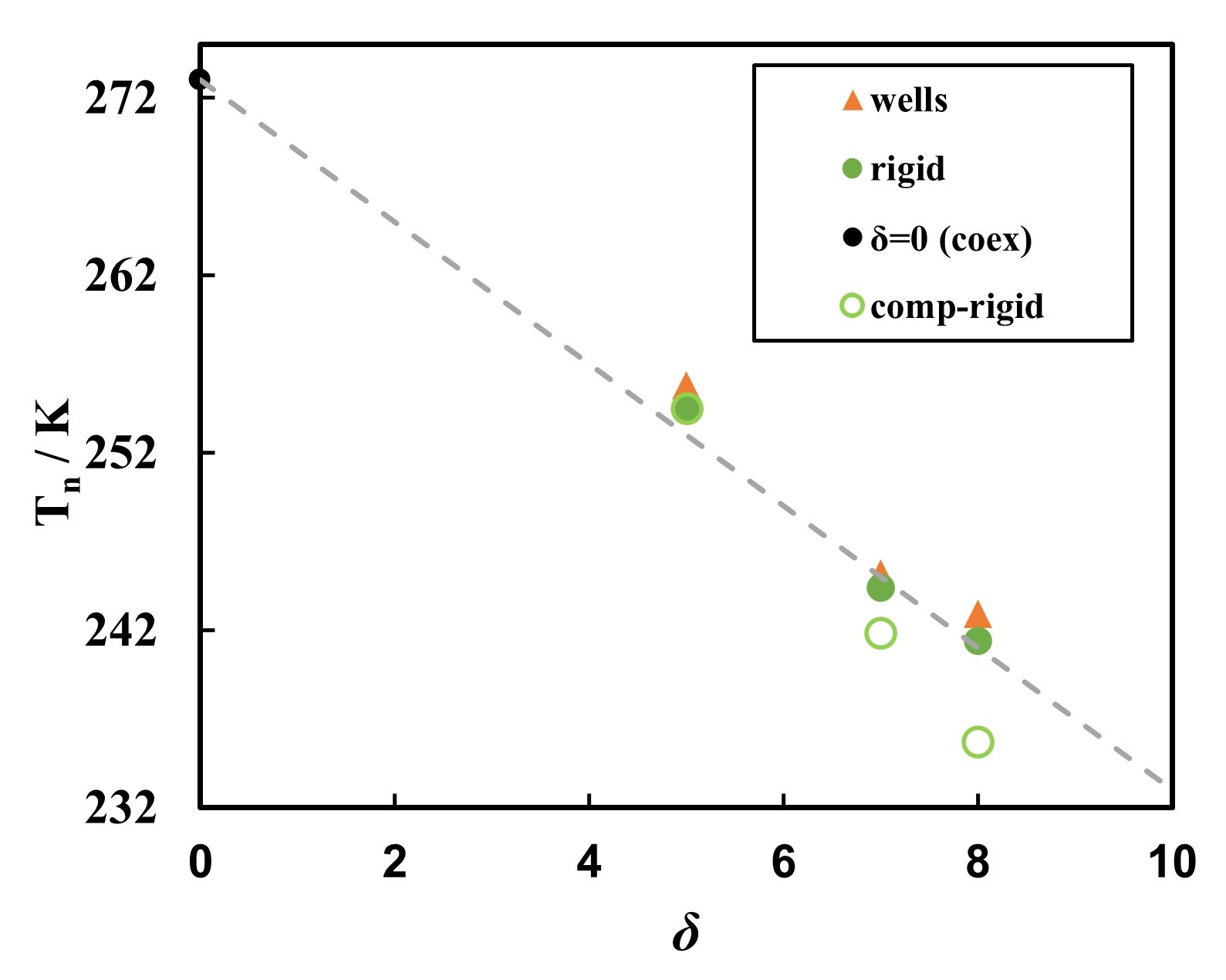}
    \caption{Nucleation temperature versus the mismatch between ice and the substrate. The two different sorts of substrates studied in this work are compared for the pII orientation. 
    The black point for $\delta=0$ corresponds to
    the ice melting temperature of the model (273 K). \textcolor{black}{Empty green circles correspond to the rigid compressed substrate}. In grey dashed, a line with -4 K/$\delta$ slope starting 
    at the coexistence point is shown for visual reference. \textcolor{black}{Symbol size coincides with that of the estimated error bar.}}
    \label{fig:nucTemp}
\end{figure}
\end{center}

In Fig. \ref{fig:nucTemp}, which is the main figure of this work, we plot T$_n$
%(defined as the temperature for which 
%a log(J/(m$^2$/s) = XX is found) 
versus the mismatch for both sorts of substrates.
Orange triangles and green dots correspond to wells and rigid substrates, respectively. 
As can be seen, as $\delta$ increases T$_n$ goes down approximately at a rate of
4 K per $\delta$ unit in both cases. In other words, if the substrate lattice is
mismatched by one per cent with respect to the ice lattice, the nucleation 
temperature goes down by about 4 K. 
This slope of 4 K/$\delta$ nicely extrapolates the nucleation data up to the coexistence
temperature (black dot in the figure) corresponding to no mismatch between ice and substrate. 
In fact, a line starting from coexistence with -4K/$\delta$ slope closely  tracks the trend of the 
calculated nucleation temperatures (grey dashed line in the figure).
Such a trend suggests a linear decrease of the nucleation temperature with the mismatch,
at least
up to the largest studied mismatch, $\delta = 8$. 
We could not tackle the study of $\delta$'s larger than 8
because liquid particles entered into the substrate.

In summary, our main conclusion is that 1 per cent of structural mismatch between the nucleating
and the substrate lattices roughly causes a 4 K decrease of the nucleation temperature. 
Given that, by construction, both lattices have the same sort of interactions, our  
study separates the effect of structural mismatch from that of inter-molecular interactions
in heterogeneous ice nucleation. 
To our knowledge, this is the first quantification of the effect of the mismatch 
alone on the ice nucleation ability of a solid substrate. Of course, the ability to 
promote nucleation in real substrates is determined by coupled 
interaction and mismatch effects \cite{fitzner2015many}. 
However, isolating the effect of
lattice mismatch may help understand and rationalize the 
ice nucleation ability of different solid substrates.

\begin{center}
\begin{figure}[!hbt] \centering
    \centering
    \includegraphics[clip,scale=0.6,angle=0.0]{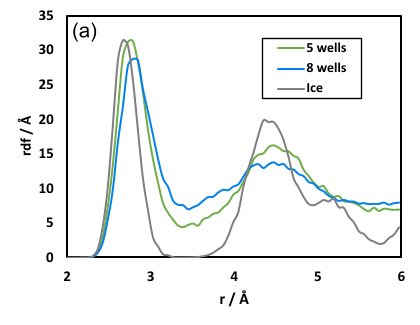}
\includegraphics[clip,scale=0.6,angle=0.0]{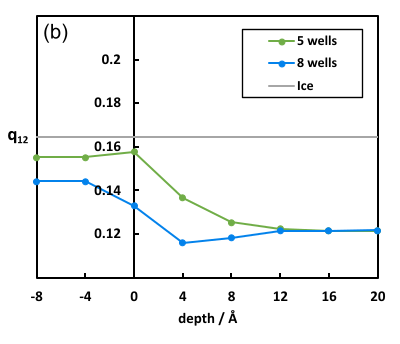}
    \caption{\textcolor{black}{(a) Rdf of the particles contained in a 7.5 \AA~ thick slab adjacent to a $\delta = 5$ and a $\delta = 8$ stretched wells substrate in green and blue respectively. In grey, the rdf of an ice slab of the same thickness. All rdf's presented correspond to simulations at 256 K. (b) q$_{12}$ profile across the interface (same color code as in (a)).}}
    \label{fig:rdfdelta}
\end{figure}
\end{center}

\begin{center}
\begin{figure}[!hbt] \centering
    \centering
    \includegraphics[clip,scale=0.65,angle=0.0]{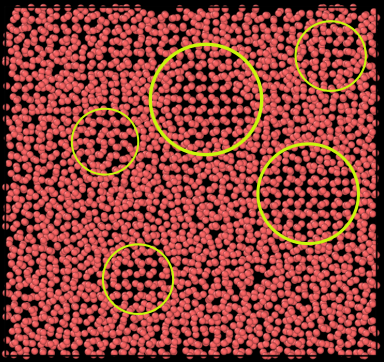}
    \caption{A 7.5 \AA~ thick liquid layer adjacent to a $\delta = 7$ wells substrate at 245.0 K. Several subcritical nuclei
    are highlighted with yellow circles.}
    \label{fig:multinucleus}
\end{figure}
\end{center}

\subsubsection{Other types of mismatch}
\textcolor{black}{In this section we discuss the impact on ice nucleation of other sorts of deformation 
of the ice lattice substrate. We first consider performing an isotropic compression of the ice lattice (as opposed to
an isotropic expansion as discussed so far). In this case, we deal with negative values of $(f-1)$, as defined 
in Eq. \ref{eq:delta}. However, according to our definition, $\delta$ is still positive because of the absolute value in 
Eq. \ref{eq:delta}. We only consider rigid substrates for this study. In Fig. \ref{fig:rate}(b) 
we compare the nucleation rate temperature dependence between compressed and expanded substrates. For a 5 per cent 
expansion/compression, J(T) is very similar between both types of deformations. However, as 
the deformation increases it seems evident that the nucleation rate is smaller for the compressed
substrate (in other words, J takes the same value at lower temperatures). The nucleation temperature corresponding to the compressed substrates is represented
as the empty green circles in Fig. \ref{fig:nucTemp}. The trend for compressed and stretched substrates is similar, although the slope is slightly larger for the former (4.5 K/$\delta$ versus 4 K/$\delta$ approximately), consistently with the fact that 
compressed substrates yield smaller rates for a given temperature. This is in 
agreement with the observation that compressed substrates interact less favourably
with water than stretched ones \cite{cox2012non}}.

\textcolor{black}{We also consider the case in which only one side of the ice unit cell is stretched. In particular, 
we build a rigid substrate by stretching 5 per cent the $z$ direction of the unit cell given in Table \ref{tab:unitcell}. 
We get
a nucleation temperature for this substrate of 266 K (7 K supercooling) by
exposing the pII orientation. This T$_n$ can be compared to the
254.6 K (18.4 K supercooling) corresponding to a substrate with a 5 per cent stretching in all directions. 
As expected, when only one dimension is stretched the nucleation temperature is higher because the substrate bears
a stronger resemblance with ice.}

\subsection{Nucleus and liquid structure}
\textcolor{black}{We focus on stretched substrates to perform
an analysis of the liquid structure adjacent to
the substrate prior to nucleation and of the 
crystal nucleus that finally grows.}
\subsubsection{Liquid structure}
\textcolor{black}{
We first examine the structure of the liquid layer adjacent to the substrate. 
In  Fig. \ref{fig:rdfdelta}(a), we compare the rdf's of the adjacent layers 
of two different wells substrates ($\delta = 5$ and $\delta = 8$) with the rdf of a bulk ice layer of the same thickness (7.5 \AA). As expected, the rdf corresponding to the layer induced by the $\delta = 5$ substrate bears more resemblance with 
the ice layer rdf, thus facilitating ice nucleation. This conclusion is confirmed
with an analysis of the q$_{12}$ 
order parameter \cite{steinhardt1983bond} computed along 4 \AA~
thick slabs parallel to the interface. 
In Fig. \ref{fig:rdfdelta}(b) we show the q$_{12}$ profiles for
the $\delta = 5$ and the $\delta = 8$ substrates in green and blue respectively. Negative 
depth values correspond to the substrate
interior and positive ones to the 
adjacent liquid. The grey horizontal
line corresponds to the bulk ice value. 
The smaller the mismatch, the closer
the q$_{12}$ profile to the ice value, 
thus confirming the expected facilitation of nucleation 
with a reduced $\delta$. For the $\delta = 5$ substrate, 
that causes nucleation at the temperature
at which these profiles have been computed (256 K), 
the structure of the liquid is affected by that
of the substrate up to a distance of 7-8 \AA~from 
the surface. For that point onward, q$_{12}$ takes 
bulk liquid values.}

\textcolor{black}{Both the rdf's  and the q$_{12}$ profiles provide an averaged structural information. 
To better understand the molecular structure of the 
interface we look at the adjacent layer of a wells substrate
with $\delta = 7$ (intermediate between those previously
analysed). 
The snapshot shown in Fig. \ref{fig:multinucleus}
reveals that the structure of the interface is
heterogeneous (the snapshot has been taken at
a temperature for which there is nucleation 
on this substrate, 245 K, but during the induction
period prior to the appearance of the critical nucleus).
One can clearly identify several sub-critical nuclei, highlighted
within yellow circles, embedded in a liquid-like
interface. 
These nuclei are likely 
responsible for the resemblance of the 
interface rdf with that of ice. 
Therefore, the interface cannot be viewed as a homogeneous wetting
layer having a diffuse ice structure, but rather as a heterogeneous 
layer alternating ice with liquid-like patches. 
A visual inspection of the interface along
the nucleation trajectory reveals that these nuclei continuously form and redissolve during the induction period until a fluctuation
leads to the formation of a nucleus big enough to proliferate (the critical cluster).} 

It is also interesting to inspect the structure across the substrate-liquid interface, before the appearance of the
crystal nucleus that eventually grows.
The green and the pink curves in Fig. \ref{fig:gr} correspond to the rdf first peak position 
across the solid-liquid interface for wells and rigid substrates, respectively. 
The peak position within the rigid substrate is fixed, whereas that in the wells substrate
gradually goes down. We discuss such variation in Sec III C, and here
we focus
on the liquid side, where there is a minimum close to the interface for both types
of substrates. The rdf peak position in the minimum (2.75 \AA) is in between that of bulk liquid (2.8 \AA),
recovered at large distance from the interface,  
and that of bulk ice (2.7 \AA). 
\textcolor{black}{We recall that 
the ice-like layer at the interface is
due to a heterogeneous distribution of
alternating
ice-like with liquid-like patches
rather than to a
homogeneous wetting of an ice-like layer (see Fig.
\ref{fig:multinucleus}).}

\subsubsection{Nucleus structure}
To gain insight on the structure of crystal nuclei we visually inspect 
clusters formed on top of both types of substrates. 
Examples of such nuclei are shown in Fig. \ref{fig:torreskio} for a given mismatch ($\delta = 7)$.
The enlargements in the right side of the figure show that both nuclei
have tilted molecular columns in the cluster peripherals.  
Recall that the substrate structure is that of a stretched ice lattice, so the tiling is likely due 
to the gradual recovery of the thermodynamically stable lattice parameters from the contact plane (where the 
cluster matches the substrate) upward (where the cluster recovers the relaxed ice structure). 

We show a more quantitative demonstration of the recovery of the ice structure along the cluster in Fig. \ref{fig:gr}, where
we plot the position of the first peak of the radial distribution function (rdf) along the direction perpendicular to 
the substrate interface. The rdf is computed along consecutive 4 \AA~ thick slabs parallel to the interface. 
Values of the x-axis equal to or smaller than 0 correspond to the substrate whereas those larger than 0 correspond to the
adjacent phase. Such phase corresponds, for the case of the orange and blue curves, to crystal nuclei as those shown in Fig. \ref{fig:torreskio}. 
The rdf first peak in these cases is computed along coin-like cylindrical slabs
whose perimeter coincides with the liquid-substrate-nucleus contact line. 
 The curves are averaged over several independent nuclei to improve statistics.
 \textcolor{black}{To select a configuration with a nucleus we first identify 
 from a potential energy drop as those shown in Fig. \ref{fig:espontaneous} the time at which
 nucleation takes place. We then look at the interface step by step starting from an 
 earlier time and 
 identify visually the emergence of the nucleus that expands and grows. We select a configuration
 where the nucleus looks as those shown Fig. \ref{fig:torreskio}: large enough to be analysed
 but not too big so that it reaches the liquid-vapor interface. Picking slightly smaller or bigger
 nuclei does not change the qualitative outcome of the analysis that follows}.
The orange curve in Fig. \ref{fig:gr} corresponds to a nucleus on a $\delta = 7$ rigid substrate. 
The rdf peak in the interior of the substrate is fixed by the lattice stretch and
is obviously larger than that of bulk ice, indicated in the figure with a horizontal grey line. 
From the substrate surface onward, the peak position gradually decreases recovering in about 
20 \AA~ the bulk ice value. This picture is fully consistent with the tilted molecular columns visible in 
Fig. \ref{fig:torreskio}.
The blue curve, corresponding to a cluster on a wells substrate with the \emph{same} mismatch ($\delta$ = 7)
shows a qualitatively similar behaviour (in the following section quantitative differences
are discussed). 
In summary, Fig. \ref{fig:gr} confirms that the tilting of the molecular columns seen in Fig. \ref{fig:snapshots}
corresponds to a gradual change from the substrate to the ice structures along the cluster.

\subsection{Effect of the lattice flexibility}
\label{sec:flex}
We have studied both rigid substrates, where particles are immobile, and 
wells substrates, where particles can move away from their lattice positions
up to 1.0215 \AA~. This enables assessing the effect of lattice
rigidity on heterogeneous ice nucleation. 
\textcolor{black}{Although our uncertainty of 0.1 in log[J/(m$^3$s)] and 0.5 K in T$_n$ brings the results of both 
types of substrate close to each other, it is clear from }
Fig. \ref{fig:rate}(a) that, for a given temperature and $\delta$, 
the nucleation rate is \textcolor{black}{systematically} lower 
for a rigid substrate. In other words, for a given rate and mismatch, the nucleation temperature is higher for the wells substrate. This difference is seen Fig. \ref{fig:nucTemp}, where the 
nucleation temperatures corresponding to the wells substrate systematically lie about 1-2 K above
those of the rigid substrate.
Therefore, \textcolor{black}{our data suggests that} lattice flexibility favours
heterogeneous ice nucleation. 

\begin{center}
\begin{figure}[!hbt] \centering
    \centering
    \includegraphics[clip,scale=0.35,angle=0.0]{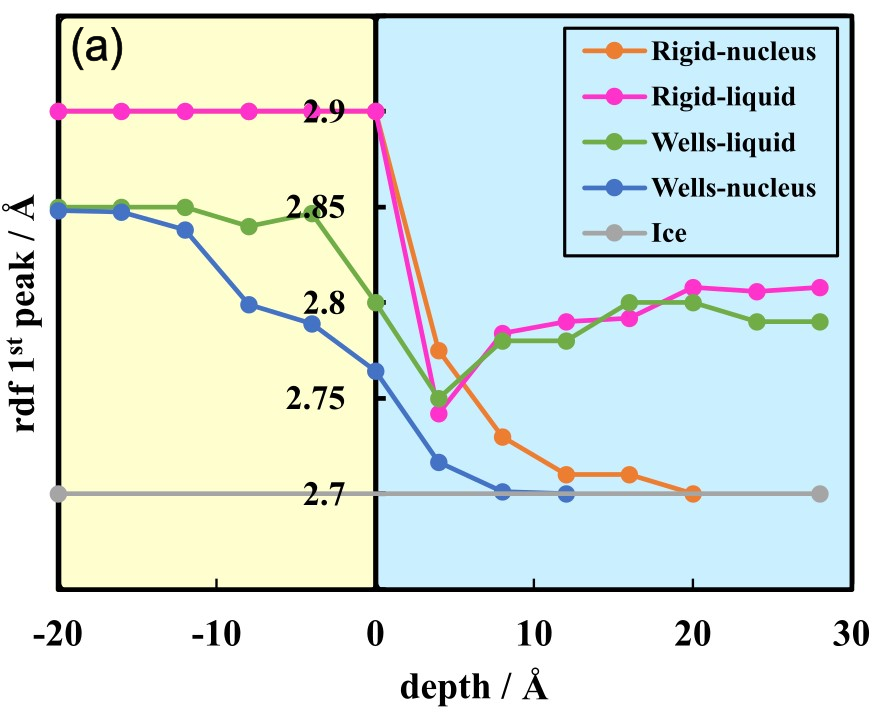}
    \includegraphics[clip,scale=0.36,angle=0.0]{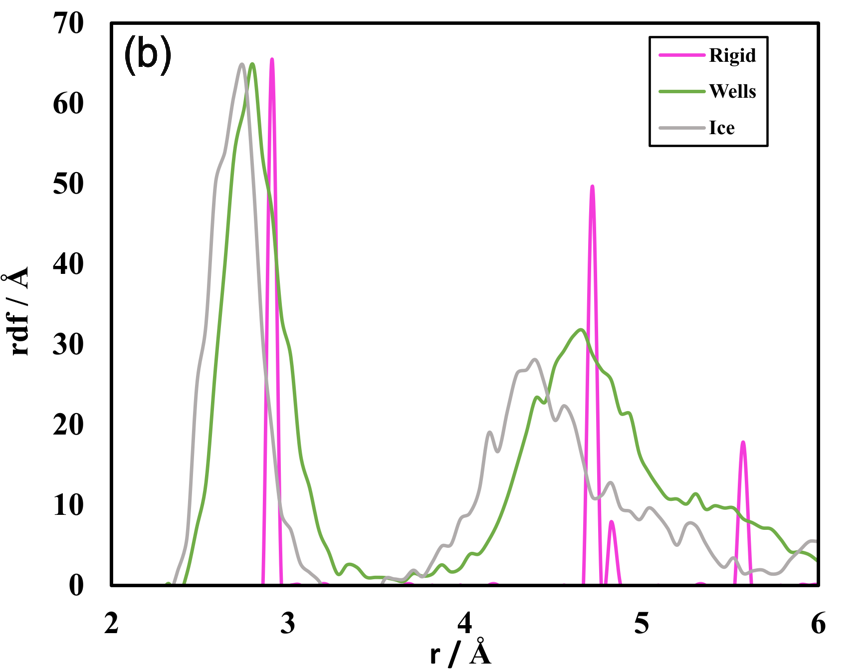}
    \caption{(a) Position of the first peak of the radial distribution 
    function along 4 \AA~ thick slabs parallel to the interface. 
Orange (blue) curve corresponds to a rigid (wells) substrate in contact with a crystal nucleus whereas
the pink (green) curve corresponds to the rigid (wells) substrate in contact with the 
liquid before the nucleus appears.
Both substrates have the same mismatch ($\delta$ = 7) and simulations are carried out at a similar temperature
(244.5 K and 245.0 K for the rigid and wells substrates, respectively).
The substrate region corresponds to depths $\le$ 0 (yellow background) and that of the adjacent phase (liquid or cluster) to
positive depths (blue background). 
The grey horizontal line indicates the value of the first peak of the radial distribution function 
for ice at 245 K. \textcolor{black}{(b) rdf of 4 \AA~ thick slabs of ice (grey) and of the $\delta = 7$ substrates mentioned in (a) in contact with the fluid (pink and green correspond to rigid
and flexible substrates as in (a)).}}
    \label{fig:gr}
\end{figure}
\end{center}

One can rationalise the better performance of the flexible substrate as an ice nucleant by looking
back at the rdf first peak position profiles shown in Fig. \ref{fig:gr}.
By comparing the blue (wells substrate-cluster) and the orange (rigid substrate-cluster) curves inside the substrate
region it can be noticed 
that, despite the fact that both substrates have the same $\delta$, the  
rdf peak in the interior of the wells substrate is closer to the
ice value.
This suggests that
particles of the flexible substrate use their freedom of motion within the wells to build a structure
in closer resemblance to that of the thermodynamically stable ice lattice. 
Such a rearrangement  
may contribute to facilitate ice nucleation. 
In any case, such an ''effective $\delta$ decrease'' is only perceivable at the level of the rdf first peak. 
The second peak already appears at the position
dictated by the imposed mismatch.
\textcolor{black}{This is shown in Fig. \ref{fig:gr}(b), where 
the following rdf's of 4 \AA~ thick slabs are compared:
ice in grey, and $\delta = 7$ rigid (flexible) 
substrate in contact with the fluid in pink (green).
As previously indicated, the first peak of the wells substrate is in between 
those of ice and the rigid substrate whereas the second peak of the wells substrate
pretty much coincides with that of the rigid substrate. Then, the effective $\delta$ reduction 
of the wells substrate means being closer to ice than the rigid substrate at the 
level of first neighbors, a parameter that has been already identified as a relevant
predictor for ice nucleating ability \cite{cox2012non}}.

\begin{center}
\begin{figure}[!hbt] \centering
    \centering
    \includegraphics[clip,scale=0.4,angle=0.0]{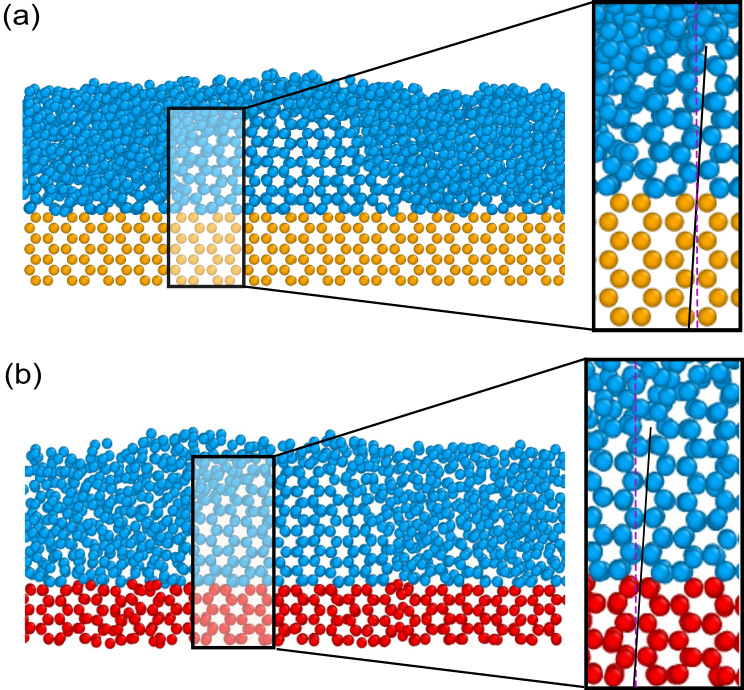}
    \caption{a): Snapshot of a slab cut through an ice cluster (in blue) nucleating on top 
    of a rigid substrate (in orange). 
    The substrate exposes the pII plane to the liquid and has a $\delta=$ 7 structural mismatch with the ice lattice.
     In the right part, a zoom view of the marked region is provided to show the tilt of the molecular columns of the cluster.
    We have included, for visual reference, dashed lines perpendicular to the substrate and
    solid lines that follow the direction of the cluster molecular columns. 
    b): Same as a) but for a wells substrate (in red).  
    }
    \label{fig:torreskio}
\end{figure}
\end{center}

Another aspect to consider is the structure of the wells substrate underneath the cluster.
A closer inspection to the enlargements provided in Fig. \ref{fig:torreskio} suggests that the tilting of the cluster molecular columns 
propagates down into the interior of the wells substrate, as if the substrate structure
adapted to that of the cluster on top. 
Again, we resort to Fig. \ref{fig:gr}
to quantify and understand this effect.
By comparing the green (wells substrate-nucleus) and the blue (wells substrate-liquid) curves in Fig. \ref{fig:gr} 
one can check if the substrate structure
changes when a nucleus appears on top. 
Clearly, in the vicinity of the interface, the blue curve is systematically below the green one, indicating that the substrate
gets more ice-like (closer to the grey curve) when having a nucleus on top. This strongly supports the view that
the substrate adapts its structure to that of the emerging nucleus.

In summary, there are two factors that may justify the better performance as an ice nucleant of the 
wells substrate: (i) the adaptation to the structure of the emerging cluster and (ii) the effective loss of 
mismatch. In any case, it is important to emphasize that flexibility 
is a second order effect. For instance, for a $\delta = 8$ mismatch, the nucleation temperature of the rigid substrate is
31.6 K below melting whereas that of the wells substrate is 30.1 K. 
Therefore, all the considerations about flexibility discussed in this section have an effect on the nucleation temperature
of less than 2 K versus the 30 K caused by mismatch.

The adaptation effect resembles the 
enhanced ice nucleating ability in  
kaolinite due to reorientation of hydroxyl groups 
 \cite{zielke2016simulations}. 
However, flexibility does not always help ice nucleation. For instance,  
in the case of substrates composed of organic molecules exposing hydroxyl groups to water, flexibility 
roughens the interface and 
is detrimental for ice nucleation \cite{qiu2017ice}.
In our case, the substrate has a flat interface and flexibility refers to the possibility of molecules
to fluctuate around their lattice positions, which enhances the nucleating ability through adaptation 
to the emerging ice structure.

\subsection{Effect of the lattice orientation}

\begin{center}
\begin{figure}[!hbt] \centering
    \centering
    \includegraphics[clip,scale=0.65,angle=0.0]{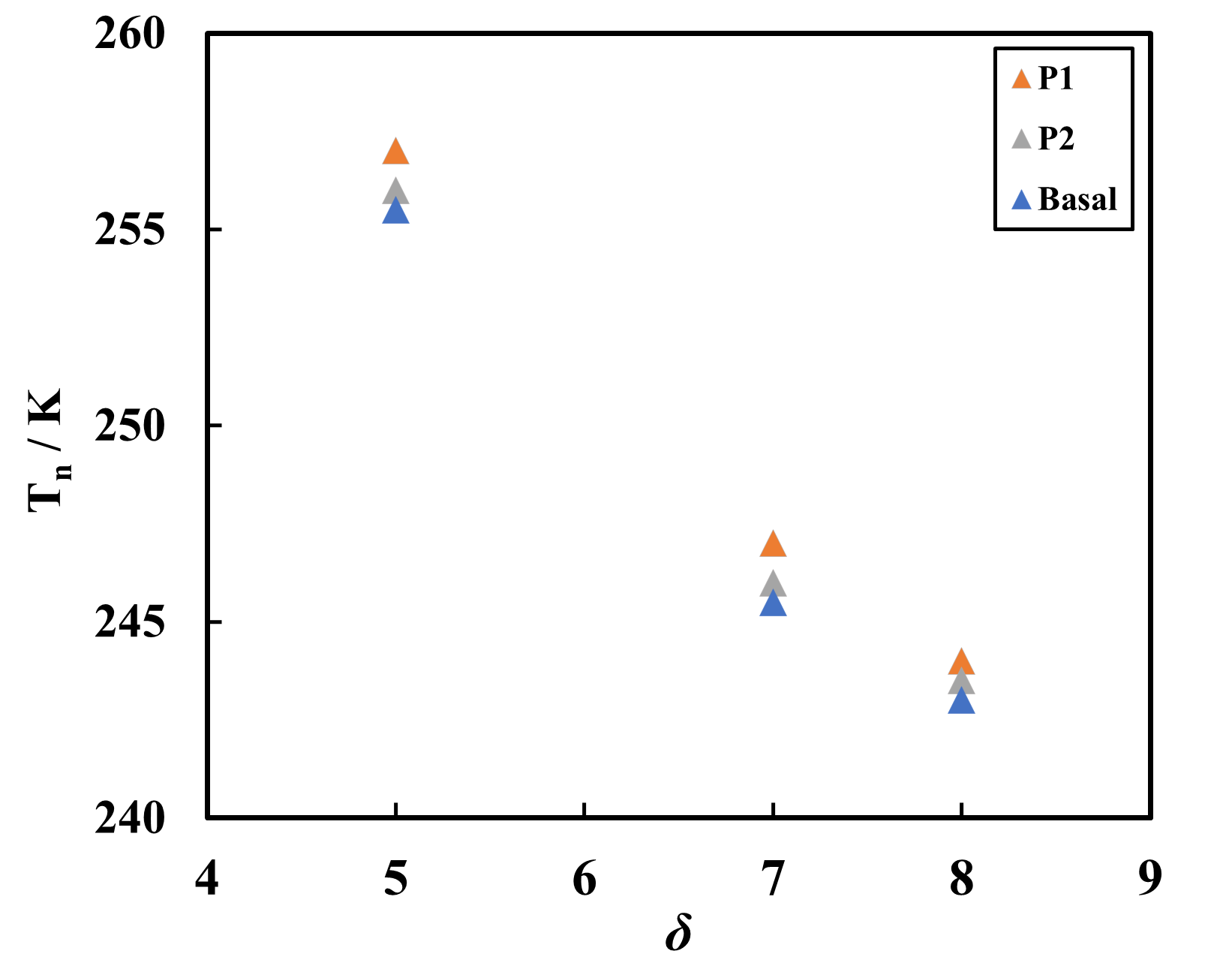}
    \caption{Nucleation temperature versus the mismatch for different orientations of stretched wells substrates.}
    \label{fig:nucTorientations}
\end{figure}
\end{center}

The results discussed so far 
correspond to the pII plane being exposed to the liquid. In this section we 
examine the ice nucleating ability 
of different orientations:
basal, pI and pII. Details on the employed system sizes to study these orientations are given in  Table \ref{tab:sistemas}. We
use wells stretched substates for this study.
We have followed the same procedure as that described for the pII plane to obtain the data shown in Fig. \ref{fig:nucTorientations}, 
where the nucleation temperature as a function of the mismatch parameter is compared for the three 
orientations of interest. 
Differences between orientations are small, but it seems that pI and basal are 
systematically  the best and
the worst ice nucleants, respectively, with a T$_n$ difference of about 1-2 K between them.
\textcolor{black}{Considering the temperature dependence of 
J shown in Fig. \ref{fig:rate}, this temperature difference means that the nucleation rate 
is 0,5-1 order of magnitude higher on the pI plane.}
The T$_n$ of the pII plane lies in between. 
\textcolor{black}{This pattern repeats itself for the three studied values of $\delta$. Thus, even though
all orientations could be considered equally good ice nucleants taking a 0.5 K error bar in T$_n$ into account, 
we consider it
is worth examining the structure of each interface in an attempt to correlate orientation with nucleation ability.}

\begin{center}
\begin{figure}[!hbt] \centering
    \centering
    \includegraphics[clip,scale=0.75,angle=0.0]{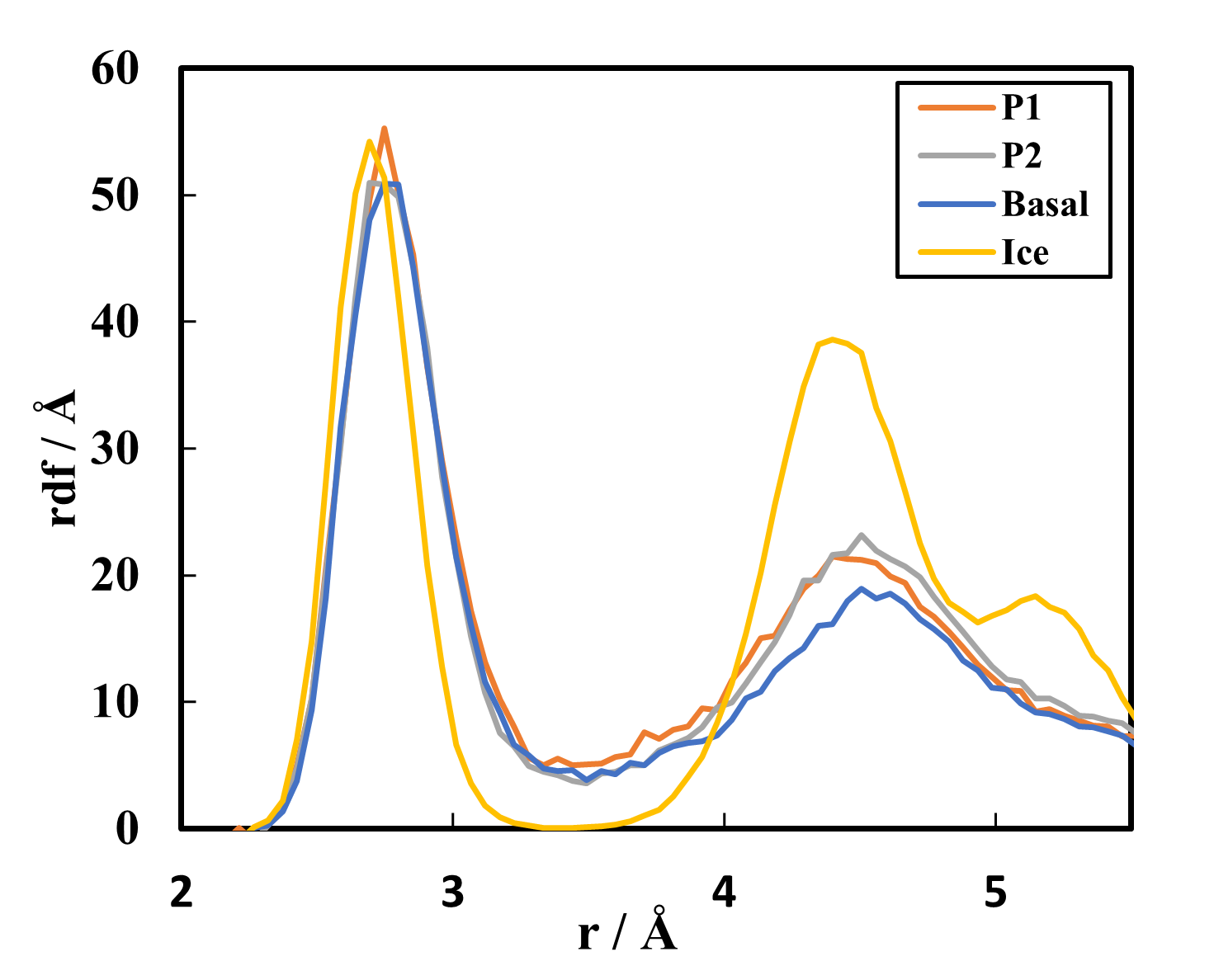}
    \caption{Radial distribution function computed in a 4 \AA~ tick liquid slab adjacent to
    a $\delta = 5$ stretched wells substrate exposing different orientations (orange, grey and
    blue curves correspond to the pI, pII and basal planes, respectively). The yellow curve
    is the bulk ice rdf. The temperature is 256 K in all cases.}
    \label{fig:gr2}
\end{figure}
\end{center}

To understand the different nucleating abilities 
of the different ice orientations 
we look at the structure of the liquid wetting each type of substrate.
In Fig. \ref{fig:gr2} we show the rdf's 
obtained in a 4 \AA~ thick slab adjacent
to $\delta = 5$ substrates at 256 K. We also include, for comparison,
the bulk ice rdf at the same temperature (multiplied by an arbitrary factor to enable
visual comparison with the un-normalised liquid slabs rdf's). 
Different orientations give rise to alike rdf's, which is unsurprising because
they also yield similar nucleation temperatures.
All orientations show a first maximum at 2.75 \AA~, at larger distances than
the first peak of ice, which is positioned at 2.70 \AA.
Small but noticeable differences can be observed, however, in the second peak. 
Clearly, the rdf second peak obtained from the liquid layer adjacent to the basal plane
is the one that lies furthest from the bulk ice second peak.
This is consistent with the fact that the basal plane is the worst nucleant.
The second peaks corresponding to the liquids adjacent to the pI and the pII orientations
are very close to each other, although it seems that the former
lies slightly closer to 
that of ice, again consistent with pI being the best nucleant.
In summary, the degree of structural similarity between the liquid adjacent to the substrate
and ice is consistent with the relative ice nucleating abilities of different orientations,
\textcolor{black}{although this should be taken with caution considering our 0.5 K error bar in T$_n$}.

\section{DISCUSSION}
\textcolor{black}{One may wonder what is the added value of using a model substrate having
the same structure and the same interactions as the ice crystal. First of all, we
are not the first ones to employ this strategy: it has been previously used for water \cite{reinhardt2014effects} (although
the effect of lattice mismatch on the nucleation temperature was not systematically quantified) and for a system of 
Lennard-Jones particles \cite{mithen2014computer,mithen2014epitaxial} (although  
the substrate-fluid interactions were weakened to prevent rapid crystallization). The 
main point of this approach is that it enables to unambiguously elucidate the role 
of lattice mismatch on heterogeneous ice nucleation, that has been long believed to be a key
factor \cite{turnbull1952nucleation}. In fact, a recent and very successful collaboration between simulations
and experiments \cite{kiselev2017active} explains the ice nucleating ability of feldspar with a
match of a ``3 × 1 array of [ice] prism face unit cells within one unit cell of the feldspar (100) surface.''
However, the validity of the lattice match as a predictor for ice nucleating ability has been challenged \cite{finnegan2003new,reinhardt2014effects,fraux2014note}.
For instance, for a simulation study where the substrate has an fcc structure, it has been concluded that $\delta =0$ is not the optimal value for nucleation \cite{fitzner2015many}.
Our approach, in apparent contrast, shows that
the best crystallizing structure corresponds to $\delta = 0$ and that the nucleation temperature
decreases \textcolor{black}{roughly} linearly with $\delta$ \textcolor{black}{(with a slightly larger slope for compressed than for expanded substrates)}. A similar conclusion, using an analogous approach, was found for the Lennard-Jones system \cite{mithen2014computer,mithen2014epitaxial} where the optimal mismatch is close to 0 (it is not strictly 0 because the crystal nucleus, due to the Laplace pressure, has a slightly different lattice parameter than the bulk solid). 
Is then the work published in Refs. \cite{mithen2014computer} and \cite{mithen2014epitaxial} and our
work in contradiction with Ref. \cite{fitzner2015many}? It is not in the sense that perhaps the best way to define structural similarity when the substrate lattice is not wurtzite-like (that of ice Ih) is not through the lattice edges, but through another structural feature that better accounts for the similarity between the ice and the substrate. In Ref. \cite{cox2012non}, for instance, a match between first neighbour distance was found to be a better 
structural predictor for ice nucleation on close packed surfaces. 
Also, in Ref. \cite{soni2021microscopic}
alternative definitions to the conventional lattice mismatch are found to better predict 
ice nucleation abilities in mineral lattices. 
Then, it may make more sense speaking of a 
\emph{structural} rather than of a \emph{lattice} mismatch. The problem is that
the definition of the former concept is not unique and requires a systematic analysis of
different sorts of substrate lattices, which is an interesting study for future work.
In our work, since we use the ice lattice as a substrate, the structural match is uniquely 
defined.}

\textcolor{black}{Another 
issue to which our approach can provide some insight is the relative 
ice-nucleating ability of the different orientations of a wurtzite-like structure (that of
ice Ih or of $\beta$-AgI, a powerful ice nucleant).
Different simulation studies on $\beta$-AgI have reported nucleation 
on the basal plane \cite{fraux2014note,zielke2015molecular}.
However, in Ref. \cite{soni2022ice} strongly different nucleating abilities between the different prism orientations
of AgI were reported. Ice nucleation was observed on the pI plane but not on the pII `` by details of the surface water
interaction''. Such details are not present in our approach because our substrate interacts
with water in the same manner as water does with itself. 
In fact, in Ref. \cite{soni2022ice} 
nucleation on the pII plane is observed when an ice-like substrate is used like in the present work. Here, we quantify the nucleating ability of the three main wurtzite orientations 
and find similar nucleation rates for all of them. 
Thereby, our work serves to \textcolor{black}{confirm}
that possible differences in nucleating abilities between the orientations of a wurtzite-type structure are not due to structural issues but rather to the specific nature
of the substrate-water interactions.} 

\textcolor{black}{In this respect, once 
again, one should reflect on the fact that the results obtained with simulations 
that aim to be realistic
may strongly vary depending on the selected force field.
This is acknowledged in Ref. \cite{sosso2016ice}, where heterogeneous nucleation on kaolinite was
studied with classical simulations: ``fully atomistic
models are needed to deal with water at complex interfaces,
such as crystalline surfaces of organic crystals or mineral dust
particles. In these cases, it remains to be seen whether the
description of the surface, and most importantly of the water surface interaction, is accurate enough to allow for reliable
results to be obtained.'' A clear example of this problem can be found in Ref. \cite{fraux2014note}, where nucleation
was studied on AgI, the following issue was encountered: ``in contrast to real AgI, the crystal began
to quickly dissolve in water, and therefore, we constrained
the Ag$^+$ and I$^−$ ions with a harmonic potential.''
Therefore, studies on heterogeneous ice
nucleation with realistic substrates, although certainly valuable and pioneer,
may draw force-field dependent conclusions. 
As an effort in improving the existing force fields to describe 
heterogeneous nucleants is under way, our work, 
even though not directly transferable to experiments due to the lack of realism 
of the substrate structure and potential, can serve to extract fundamental conclusions
about structural effects on ice nucleating ability.} 

\textcolor{black}{In a context where we do not have 
tested surface-water interactions, where
the role of lattice mismatch is debated and where the nucleating ability of the
main wurtzite orientations needs to be further examined, we believe it is worth 
pursuing a simple approach
like ours, where the ambiguity in defining structural matching and the 
coupling between structure and interactions have been removed. 
In Section \ref{sec:concl} we summarize the main conclusions we draw from our work. 
The added value of this paper is, therefore, 
having devised and applied a strategy to quantify and isolate the role of lattice mismatch on 
heterogeneous ice nucleation and to examine 
the effect of wurtzite lattice orientation and flexibility. 
We acknowledge that our strategy does not enable a direct 
comparison with experiments, but this is also a problem 
for simulations that rely on the performance of force fields that have not been
thoroughly tested. In order to test a force field it is necessary to check if it gives
predictions that are close to experimental measurements. 
A property that can be computed in simulations and measured in experiments is the nucleation rate, that has been 
successfully compared for the case of homogeneous ice nucleation \cite{espinosa2016interfacial,espinosa2018homogeneous}.
However, to our knowledge, there is no
direct comparison between experimental and simulated heterogeneous nucleation rates to date 
and further work is required to achieve such 
milestone.}

\textcolor{black}{
In immersion freezing experiments 
freezing takes place in an ensemble of supercooled water drops containing 
ice nucleating particles. 
The heterogeneous nucleation 
rate can be inferred by monitoring the fraction of frozen drops along time \cite{knopf2020stochastic,cornwell2021development}. Through the nucleation rate,
a direct comparison between simulations and real measurements could be established. 
This sort of  comparison has been successfully done so far for homogeneous nucleation 
in supercooled water \cite{espinosa2018homogeneous}. In contrast, we are unaware of the existence of a 
similar comparison for heterogeneous nucleation. There are several reasons for this 
lack of simulation-experimental bridges in heterogeneous ice nucleation. 
On the one hand,
it is only recently that heterogeneous nucleation rates are being experimentally 
determined (traditionally, n$_s$, the number of
active sites per particle, has been used to characterise heterogeneous nucleation \cite{murray2012ice,knopf2020stochastic}, but this parameter does not take into account the stochastic and time dependent character 
of nucleation \cite{knopf2020stochastic}).
On the simulation side, the main drawback is the lack of reliable substrate-water interactions
that enable for quantitative predictions of the nucleation rate. \textcolor{black}{Moreover, running simulations with atomistic models is much more complex than with the mW model.}
Even though it is not the purpose of this work to establish a direct comparison with 
experiments (we have used a model rather than a real susbstrate with
the aim of investigating the influence of the substrate structure on ice nucleation), 
we can at least compare the order of magnitude of our rates with typical experimental values. 
In Ref. \cite{cornwell2021development}, 
the nucleation rate reported for dust ice nucleating particles at 30 K supercooling
is of the order of 10$^{10}$m$^{-2}$s$^{-1}$. At this supercooling, the nucleation 
rate of the least efficient simulated substrate, the $\delta = 8$ rigid one, is of the order
of 10$^{24}$m$^{-2}$s$^{-1}$. There is a huge difference in orders of magnitude that
goes in the expected direction: Obviously, 
an ice-like substrate with water-like interactions like ours, is much more efficient
in nucleating ice than a dust mineral that bears much less resemblance with ice. 
A great effort is needed, both from simulations and experiments, to get 
heterogeneous
nucleation data that are consistent with each other. 
It would be highly desirable to achieve such consistence because it would 
greatly enhance our confidence in simulation predictions regarding the
structure of the substrate-water interface. In the mean time, a study like 
the present one that focuses on understanding how the structure of a model substrate
affects its nucleation efficiency can provide insights on the problem of 
heterogeneous ice nucleation.}

\textcolor{black}{We stress that we do not want our work to convey the message that lattice structure is the only important factor in heterogeneous ice nucleation. We have simply focused on this factor alone because we believe that when there are many variables affecting a complex phenomenon it is of interest to analyse them one by one. The message that, when all other factors are removed, 
lattice mismatch is important, does not contradict that such parameter may not be a good descriptor
when coupled structural and interaction effects play a simultaneous role \cite{bi2017enhanced,fitzner2015many,glatz2018heterogeneous,cox2015molecularI,cox2015molecularII,valeriani2022deep,soni2021microscopic,reinhardt2014effects,li2017roles,zhang2018control,lupi2014does,lu2021effect,cox2013microscopic,fitzner2020predicting}.} 

%\textcolor{black}{
%Finally, it is interesting to realise how much the substrate accelerates nucleation by comparing 
%homogeneous to heterogeneous nucleation times. 
%To establish this comparison we chose the $\delta =  8$ wells substrate at 30 K supercooling.
%The corresponding heterogeneous nucleation rate interpolated from Fig. \ref{fig:rate}
%is 10$^{-23,5}$ m$^{-2}$ s$^{-1}$.
%This leads to a nucleation time on the area of our substrate (145 nm$^2$) of $\sim$ 10 ns.
%At the same supercooling, the homogeneous nucleation rate for the mW 
%water model can be interpolated from Ref. \cite{espinosa2016seeding}: J$_{hom}$ = %10$^{-25}$ m$^{-3}$s$^{-1}$. For a system like ours containing 30600 molecules
%this gives a nucleation time of the order of 10$^{58}$ ns. This comparison
%illustrates how using stretched ice-like
%substrates enormously decreases the nucleation time.}

\section{\label{sec:concl}CONCLUSIONS}
We have performed molecular simulations of supercooled water in contact with substrates composed
of water molecules 
having a stretched/compressed ice structure. This strategy enables us to establish a direct relationship
between the ice nucleating ability (quantified by means of the temperature
at which nucleation is observed) and the mismatch between the ice and the substrate
structures. These are the main conclusions we draw from our work:
\begin{itemize}
    \item A one per cent increase of mismatch between a stretched ice lattice and ice itself decreases the nucleation temperature by approximately 4 K.
    \item \textcolor{black}{Stretching causes a similar
    effect as compressing for low-moderate deformations. For high deformations compressing 
    leads to poorer nucleants.} 
    \item The nucleating abilities of the main ice faces (basal, pI and pII) 
    are very similar to each other, 
    being the pI orientation the most efficient with a nucleation 
    temperature about 1-2 K above that of 
    the basal plane, which is the poorest nucleant. 
    \item Lattice flexibility of the substrate enhances ice nucleation through both an effective loss of mismatch and an adaptation 
    of the substrate structure to that of the emerging nucleus.
    \item The structure of the crystal nucleus matches that of the substrate in the contact
    plane and gradually recovers that of bulk ice in more distant planes.
    \item Sub critical clusters are formed and redissolved on the liquid-substrate
    contact plane during the induction period prior to the formation of the critical nucleus.
\end{itemize}
We have used a simplified monoatomic water model to perform this work. It would be interesting
to test in the future whether a similar behaviour is recovered with more realistic potentials. 
One can also study with this approach the role of substrate-water interactions on 
heterogeneous ice nucleation
or even explore substrate structures inspired in real nucleants like inorganic minerals. 

 \section{Data Availability Statement}
The data that supports the findings of this study are available within the article.

\begin{acknowledgments}
This work was funded by Grants No. PID2019-105898GB-C21, PID2019-105898GA-C22, PID2022-136919NB-C31 and PID2022-136919NB-C32 of the MICINN. The authors gratefully acknowledge the Universidad Politecnica de Madrid (www.upm.es) for providing computing resources on Magerit Supercomputer.
M.M.C. and J.R. acknowledge CAM and UPM for financial support of this work through the CavItieS project No. APOYO-JOVENES-01HQ1S-129-B5E4MM from ``Accion financiada por la Comunidad de Madrid en el marco del Convenio Plurianual con la Universidad Politecnica de Madrid en la linea de actuacion estimulo a la investigacion de jovenes doctores'' and CAM under the Multiannual Agreement with UPM in the line Excellence Programme for University Professors, in the context of the V PRICIT (Regional Programme of Research and Technological Innovation). 
\end{acknowledgments}

\section{REFERENCES}

\end{document}